\input{psfig.sty}
\documentstyle[12pt]{article}
\newcommand{\journal}[4]{{\em #1~}#2\,(19#3)\,#4;}

\newcommand{\pr}{\journal {Phys. Rev.}}

\newcommand{\cmp}{\journal {Comm. Math. Phys.}}

\newcommand{\np}{\journal {Nucl. Phys.}} 
\newcommand{\pl}{\journal {Phys. Lett.}}

\makeatletter 
\@addtoreset{equation}{section} 
\makeatother 
 
\renewcommand{\a}{\alpha} 
\renewcommand{\b}{\beta} 
\renewcommand{\d}{\delta}  
\newcommand{\e}{\varepsilon} 
\newcommand{\f}{\phi} 
\newcommand{\g}{\gamma} \newcommand{\G}{\Gamma} 
 
\renewcommand{\L}{\Lambda}
\renewcommand{\l}{\lambda} 
 
\newcommand{\m}{\mu} 
\newcommand{\n}{\nu} 
\newcommand{\mn}{{\m\n}} 
\renewcommand{\o}{\omega}  
\newcommand{\p}{\psi} 
\newcommand{\pb}{\bar\psi} 
\newcommand{\r}{\rho}

\newcommand{\s}{\sigma} \renewcommand{\S}{\Sigma}

\newcommand{\DD}{{\cal D}} 
 
\newcommand{\FF}{{\cal F}}

\newcommand{\NN}{{\cal N}} 
\newcommand{\OO}{{\cal O}}

\newcommand{\TT}{{\cal T}}

\newcommand{\complex}{{\kern .1em {\raise .47ex 
\hbox {$\scriptscriptstyle |$}} 
\kern -.4em {\rm C}}} 
\newcommand{\real}{{{\rm I} \kern -.19em {\rm R}}} 
\newcommand{\rational}{{\kern .1em {\raise .47ex 
\hbox{$\scripscriptstyle |$}} 
\kern -.35em {\rm Q}}} 
\renewcommand{\natural}{{\vrule height 1.6ex width 
.05em depth 0ex \kern -.35em {\rm N}}}
 
\newcommand{\ipr}{\!\cdot\!}

\newcommand{\tr}{{\rm {Tr} \,}} 
 
\renewcommand{\exp}{{\rm \ {exp}\,}} 
\newcommand{\cb}{{\bar c}}

\newcommand{\gh}{{\hat g}} 
\newcommand{\half}{\frac 1 2} 
 
\newcommand{\pa}{\partial} 
\newcommand{\pad}[2]{{\frac{\partial #1}{\partial #2}}}

\newcommand{\sla}{\raise.15ex\hbox{$/$}\kern -.8em} 
 
\newcommand{\twiddle}{\lower.9ex\rlap{$\kern -.1em\scriptstyle\sim$}}

\newcommand{\vev}[1]{\left\langle {#1}\right\rangle}

\newcommand{\equ}[1]{~(\ref{#1})}

\newcommand{\eq}{\begin{equation}} 
\newcommand{\eqn}[1]{\label{#1}\end{equation}} 
\newcommand{\eea}{\end{eqnarray}} 
\newcommand{\eqa}{\begin{eqnarray}}
\newcommand{\eqan}[1]{\label{#1}\end{eqnarray}} 
\newcommand{\ba}[1]{\begin{equation}\begin{array}{#1}} 
\newcommand{\ea}[1]{\end{array}\label{#1}\end{equation}} 
\newcommand{\eqac}{\begin{equation}\begin{array}{rcl}} 
\newcommand{\eqacn}[1]{\end{array}\label{#1}\end{equation}}

\renewcommand{\pad}[2]{{\displaystyle{\frac{\partial #1}{\partial #2}}}}

\newcommand{\one}{{\bf 1}} 
\renewcommand{\sb}{{\bar \s}} 
\newcommand{\gb}{{\bar \g}}
\newcommand{\GeV}{{\rm GeV}} 
\newcommand{\MeV}{{\rm MeV}} 
\newcommand{\fm}{{\rm fm}}
 
\begin{document} 
\def\ftoday{{\sl  \number\day \space\ifcase\month  
\or Janvier\or F\'evrier\or Mars\or avril\or Mai 
\or Juin\or Juillet\or Ao\^ut\or Septembre\or Octobre 
\or Novembre \or D\'ecembre\fi 
\space  \number\year}}     
\titlepage 
%
{
\begin{center} 
{ \huge  The vacuum energy density of QCD  with $n_f$=3
Quark Flavors}  
\vspace{2ex} 
 
{\Large Martin Schaden\footnote{e-mail~: ms68@scires.nyu.edu\\ 
\indent{~~offers} to continue this research are welcome.}\\  
{\it\large Physics Department, New York University,\\ 4 Washington Place, 
New York, N.Y. 10003}} 
\end{center} 
\vspace{4ex} 
 
\begin{center} 
\bf ABSTRACT 
\end{center} 
 
An equivariant BRST-construction is used  to 
define the continuum $SU(3)$ gauge theory on a finite torus. I
corroborate previous results using renormalization group techniques by
explicitly computing the measure on the moduli-space of the model with
$3$ quark flavors  to two loops. I find that
the correction to the maximum of the one-loop effective action is indeed of
order $g^2$ in the critical covariant gauge. The leading logarithmic
corrections  from higher loops are
also shown to be suppressed by at least one order of $g^2$. I
therefore can relate the expectation value $\kappa^4=-\vev{\tr \gb^2
}$ of the moduli $\gb$ to the asymptotic scale parameter of the  
modified minimal subtraction scheme. An immediate consequence is the
non-perturbative result  $\vev{\Theta_{\m\m}}=-\frac{3}{8\pi^2}\kappa^4
= - \frac{(3\exp 2)}{4^{5/3}\pi^2}
\L^4_{\overline{MS}}= -0.2228\dots\L^4_{\overline{MS}}$
for the expectation value of the trace of the energy momentum tensor
of QCD with three quark flavors. This relation compares favorably  with
phenomenological estimates of 
$\vev{\Theta_{\m\m}}$ from QCD sumrules for the charmonium system and
$\L^{(3)}_{\overline{MS}}$ from  $\tau$-decay. 

PACS: 11.15.-q 11.15.Bt 11.15.Tk 11.15.Bx\hfill\break 
NYU-TH--98/02/22\hfil\hfil February 1998 
\vfill 
 
\newpage 
\def\be{\begin{eqnarray}} 
\def\ee{\end{eqnarray}} 
\def\nn{\nonumber} 
\section{Introduction}
The perturbative regime of ordinary SU(n) gauge theory with $n_f$
quarks in the fundamental representation of the group has been
thoroughly investigated in covariant gauges. The ultraviolet
divergences of the loop expansion can be regularized by analytic
continuation in the dimension of Euclidean space-time. The
infrared divergences of ordinary covariant perturbation theory 
due to the massless gauge bosons and ghosts are however 
an indication of the limited  validity of this asymptotic expansion.
The infrared problem is partially circumvented by the Operator Product
Expansion (OPE)\cite{wi69} which is parametrized by non-perturbative
matrix elements of certain local operators.  Values for  these matrix
elements generally  have to be extracted from 
experiment or from lattice simulations. I will here calculate one such
non-perturbative quantity, the vacuum expectation value of the trace
of the energy momentum tensor. The computation of this expectation
value in terms of the asymptotic scale parameter requires
little more than an asymptotic analysis of the theory.   

The basic idea for this calculation was presented in
ref.$II$\cite{II}. One uses the fact
that infrared divergences are regulated in a gauge invariant
fashion by considering the theory on a compact space-time manifold such as a
torus. Asymptotic freedom asserts that
perturbation theory is  reasonably accurate for sufficiently small
physical volume of the compact Euclidean manifold and one computes the
deviations that arise as the volume of the torus becomes large\cite{II,I}.

The main challenge is to define the model on a
compact Euclidean space-time in a covariant manner. This separates (dynamical)
degrees of freedom that propagate from those which do not (moduli).  
In a reasonably gauge-fixed theory on a torus, the latter are
a finite set of parameters and the loop
expansion for the {\it dynamical} degrees of freedom depends on their values.
In {\it covariant} gauges, that is gauges which 
preserve {\it all} the isometries of the compact space-time, the
moduli  are generally space-time independent constants which are
related to vacuum expectation values of the basic
fields and their composites. They differ from (dimensionful) couplings
in that they take values in a certain (moduli) space with a
measure which has to be calculated.  The
volume dependence of this measure essentially  determines the
transition to the thermodynamic limit of the theory: at sufficiently
small space-time volumes the measure on the moduli space is
almost flat but it constrains the relevant moduli-space to a
submanifold in the thermodynamic limit that is characterized by a few
constants. 

The problem of constructing a sensible covariantly gauge-fixed $SU(n)$ theory
with quarks in the fundamental representation in a
compact Euclidean space-time was solved in ref.$I$\cite{I} using an
equivariant BRST-quantization that eliminates the constant zero-modes of
the Faddeev-Popov ghosts. The argument for this construction can be
summarized as follows. With fermions in the fundamental
representation of the gauge group on a torus, the
 gauge field $A_\mu$ and Faddeev-Popov ghosts in covariant gauges
 are periodic fields\cite{tH79}. Global gauge invariance of the action implies
that the constant (anti-) ghost in ordinary covariant gauges
decouples. One can show that this leads to a vanishing
partition function of such a continuum model. [One
verifies\cite{II,brs96} that the Witten-index vanishes when the gauge
fixing is viewed  as a  
Topological Quantum Field Theory (TQFT) on the gauge group]. 
The equivariant BRST-quantization eliminates this problem without
destroying the covariance of the gauge-fixed model. Generic zero-modes
of the ghosts in this case are absent for the finite torus.
The construction introduces constant ghosts with ghost number $0,\pm
1$ and $\pm 2$ that form  part of the moduli space. There are no
zero-modes of the antiperiodic fermions to contend with and the 
remaining zero-modes of the gauge field are bosonic and do not give 
a vanishing partition 
function. These zero modes are addititonal moduli, but as observed
in\cite{Hosotani}, zero-modes of the gauge field on a torus with
fermions in the fundamental representation are innocuous and can be
neglected for sufficiently large volumes. 

Let me also justify  considering the theory on a symmetric torus
instead of some other compact manifold.  The predominant
reason is that this manifold  preserves translational
invariance and that the thermodynamic limit of interest is thus
probably approached smoothly and uniformly. A torus of finite volume
breaks the rotational invariance of Euclidean space, but
this symmetry is not associated with any scale and generally believed
to be recovered in the thermodynamic limit. A torus with {\it
periodic} boundary conditions for the gauge fields also admits {\it
anti-periodic} fermions in the fundamental representation, i.e. quarks.
In addition, the  space of gauge orbits on any finite torus with
periodic boundary conditions is connected and the gauge dependence
of topological sectors\cite{brs96} is therefore absent. Because the orbit
space is connected, there is no strong CP-violation on
any finite torus with periodic boundary conditions for the gauge
fields. Instantons and anti-instantons appear (and disappear) pairwise
and the Pontryagin number vanishes. Due to
translational invariance there is some hope that the net effect of
such meta-stable configurations can be summarily described by  constant
moduli. We will see that the moduli in fact saturate the vacuum energy
density {\it completely} and that there are no 
``other'' non-perturbative contributions to this quantity. A
resolution of the $U_A(1)$-problem in this context is on the other
hand still outstanding. [The solution proposed by
Veneziano\cite{ve79} appears promising if the axial vector
ghost has an effective  mass that is inversely related to the volume
of the torus. The susceptibility would vanish for any
finite torus but the $U_A(1)$ problem could nevertheless be solved in the
thermodynamic limit where this (conjectured) ghost becomes
massless. Such an axial (Goldstone) vector ghost has however not been
identified.]  Let me note here that the  $U_A(1)$ ``problem''
is peculiar to {\it covariant} gauges, since 
its formulation requires the use of Goldstone's theorem to saturate
Ward identities  of the conserved, but gauge dependent, anomalous $U_A(1)$
current. 

A technical argument for using a symmetric torus as compact Euclidean
space-time is that perturbation theory on this manifold is relatively 
straightforward and that the thermodynamic limit of the expressions
is readily taken.  The mode expansion can be explicitly performed and
an analytic continuation in the dimension $D$ of the torus regulates the
UV-divergent mode-sums of the loop expansion as it regulates the
loop-integrals in ordinary (covariant) perturbation theory. 

In $II$ the model proposed in $I$ was qualitatively analyzed in
detail. It was argued that perturbation theory {\it for gauge
dependent} quantities is reliable in the vicinity of fixed points
of the parameter {\it and} moduli space only. Technically one
observes that ``dangerous'' leading log corrections which tend to
invalidate the perturbative expansion are absent at certain points of
the parameter and moduli space. As we will see explicitly, the problem
in perturbative QCD generally is that some  of the leading log
corrections of the loop expansion arise because generic gauge
parameters and moduli flow to certain nontrivial fixed points.  Using
the renormalization group (RG) equation, these corrections can in
principle be resummed (in practice this is approximately possible only
if the gauge parameters and moduli are already  sufficiently close to
a fixed point), leading to a nonanalytic dependence on the coupling of
the gauge parameters and 
moduli.  The mess can be partially avoided by perturbatively expanding
at the fixed points for these parameters from the outset. The remaining
non-analytic dependence of the perturbative expansion is then due to
the scale dependence of the loop expansion parameter $g$,
asymptotically described by $\L_{ASP}$,  the only physical parameter of
the theory.  
 
There is another (non-perturbative) argument for expanding at the
fixed points of gauge parameters and moduli. For generic values of the
parameters and moduli, the loop expansion is
{\it sensitive} to small changes in their values. One may imagine that
such a change could effectively be induced by taking certain
non-perturbative configurations into 
account. An example in point is the scale dependence of the covariant
gauge parameter $\alpha$. Non-perturbative arguments\cite{I} suggest
that $\alpha$ is scale independent, since it is a coupling parameter
of a TQFT. In a perturbative calculation, $\alpha$ does, however,
generically show an asymptotic scale dependence.  It 
is suggestive to consider this apparent scale dependence as 
due to Gribov copies of (perturbative) configurations that have been
{\it neglected} in the loop expansion. At the fixed point, the combined
effect of the neglected Gribov copies is at least asymptotically 
{\it not} scale dependent. Since this scale dependence of $\alpha$ is
absent {\it for any} Green function, one may argue
that the influence of other Gribov regions cancels at
the fixed point and that the perturbative answer in this case is
asymptotically correct. A similar argument can be used to infer
that contributions from {\it neglected} non-perturbative
configurations do not change the value of the moduli at fixed
points. A perturbative evaluation of the model is thus
asymptotically {\it self-consistent} only in the vicinity of a fixed point in
the parameter and moduli space. The one-loop calculations in $II$
revealed  nontrivial fixed points in the moduli and parameter
space of that model for $n_f\leq n$, and the corresponding gauges were
called Critical Covariant Gauges (CCG). The 1-loop effective potential
for the moduli implied that the global $SU(n)$ symmetry of the
gauge-fixed theory is spontaneously broken to
$U(1)^{n-1}$ in CCG and the (planar) contribution to the Wilson
loop from the corresponding Goldstone bosons showed confining behavior.

In section~2 of this article I give a concise description of the $SU(n)$-model
discussed in $II$ and also present a possible extension for
$SU(n>2)$. The action and moduli space one has to consider simplifies
considerably in the special case $n_f=n$. The
CCG with $\d=1,\, \a=3$ for $n_f=n$ is a subset of the class of
gauges with $\d=1$ in which some of the moduli decouple. The
simplified action we will subsequently  consider is given
by\equ{effaction0}. In section~3 the two-loop effective action for the
remaining moduli is calculated  to two loops. It is expressed in terms of
renormalized quantities in section~4 where I also argue that higher
order corrections are suppressed in the CCG with $\a=3$.          
In section~5 the scale $\kappa$ of the spontaneously
broken $SU(n)$ symmetry is finally related to  $\L_{\overline{MS}}$ and
section~6 uses this result to obtain the expectation value of the trace of
the energy momentum tensor in terms of $\L_{\overline{MS}}$. I conclude
by comparing with the phenomenology of QCD sumrules.

\section{The Model}   
I will consider a special case of the covariant model of Euclidean QCD
on a finite D-dimensional symmetric hypertorus $\TT_D=L\times
L\times\dots\times L$ proposed in $II$.  The boundary conditions for the gauge
field $A_\m(x)$ and Faddeev-Popov ghosts $c(x)$ and $\cb(x)$ are
periodic and the model is descibed by the tree-level action  
\be\label{effaction} 
S_0&=&S_C + 2\int_{\TT_D} dx\ \tr\left[\frac{1}{2\a}(\pa\ipr A(x))^2 + 
\d\cb(x)\, D^A\ipr\pa c(x) + (1-\d)\cb(x)\pa\ipr D^A c(x) + 
\right.\nn\\ 
&&\qquad +\a\d(1-\d)g^2\cb(x)\cb(x) c(x) c(x)-\d [\cb(x),c(x)]\gb 
+\frac{1}{2\a g^2}\gb^2 -\frac{1}{\a g}\gb \pa\ipr A -\nn\\ 
&&\qquad\left. - \s c(x) c(x)-\a(1-\d)\f\cb(x)\cb(x) 
-\frac{1}{g^2}\s\f +\frac{1}{g} \g\cb(x) + \frac{1}{g}\sb c(x) \right], 
\ee 
where $S_C$ is the Yang-Mills action of an $SU(n)$ gauge theory with 
$n_f$ quark flavors in the fundamental representation   
\eq 
S_C = \int_{\TT_D} \half \tr F_\mn(A) F_\mn(A)+ \sum_{i=1}^{n_f} \pb_i 
(\sla D -m_i)\p_i \ . 
\eqn{classaction} 
The Euclidean Dirac operator here is\footnote{The Euclidean Dirac 
matrices $\g_\m$ satisfy  $\g_\m\g_\n +\g_\n\g_\m =2\d_\mn \one$. I will generally suppress color indices for notational
clarity. Except for $\p_i,\pb_i$, vectors of   
the fundamental representations of  $SU(n)$,  all fields are 
traceless, anti-hermitian $n\times n$ matrices in this notation  and 
transform under the adjoint representation of the group. The
commutator $[\cdot,\cdot]$ is graded by the ghost number.} 
\eq 
\sla D\p_i=\g_\m (\pa_\m +g A_\m)\p_i 
\eqn{dirac} 
and  the field strength $F_\mn(A)$ is
\eq 
F_\mn(A) = \pa_\m A_\n-\pa_\n A_\m +g[A_\m, A_\n] 
\eqn{fieldstrength} 
Note that a CP-violating term proportional to the Pontryagin number 
$\int_\TT \tr F\wedge F$ vanishes on a torus with periodic boundary
conditions\footnote{``Toron'' sectors with
non-vanishing Pontryagin number exist on a torus with
twisted boundary conditions. These however can only be imposed in the
absence of quarks in the fundamental representation\cite{tH79}.}. I therefore
did not include such a term in\equ{classaction}. As discussed in the
introduction, there is no strong CP violation on a {\it finite} torus
with periodic gauge fields. 

The action\equ{effaction} is on-shell invariant under the equivariant
BRST-symmetry generated by the operator $s$. Its action on the fields
is  
\be\label{sdef} 
s A_\mu(x) &=& D^{A }_\mu c(x)-[\o, A _\mu(x)]\nn\\ 
s c(x) &=& -[\o, c(x)] -\frac{g}{2}[ c (x), c (x)] -\frac{1}{g}\f\nn\\ 
s \o &=& -\half [ \o,\o] +\f\nn\\ 
s \f &=& -[\o, \f]\nn\\ 
s \cb(x)&=&-[\o,\cb(x)] + b(x)\nn\\ 
s b(x) &=& -[\o, b(x)] + [\f, \cb(x)]\nn\\ 
s \s &=& -[\o,\s]  +\sb \nn\\ 
s \sb &=& -[\o, \sb] + [\f, \s]\nn\\ 
s \gb &=& -[\o,\gb] +\g\nn\\ 
s \g &=& -[\o, \g] + [\f, \gb]\nn\\ 
s \p_i(x) &=&-\o\p_i(x) -g c(x)\p_i(x) \nn\\  
s \pb_i(x) &=& -\pb_i(x)\o -g\pb_i(x) c(x) 
\ee 
where 
\eq 
D^A_\m c(x) = \pa_\m c(x) + g[A_\m(x), c(x)] 
\eqn{covderivative} 
is the usual covariant derivative in the adjoint representation. In
\equ{effaction} the Nakanishi-Lautrup field $b(x)$ has been eliminated
using the equation of motion
\eq 
b(x)=(\gb/g -\pa\ipr A(x))/\a -\d g[\cb(x), c(x)]. 
\eqn{EMb}   

It is straightforward to show that the BRST-operator defined above is  
nilpotent  on any element of the graded algebra constructed from the 
fields of Table~1: 
\eq  
s^2=0 
\eqn{nilpotency} 
\begin{center} 
\begin{tabular}{|l|r|r||r|r|r|r|r|r|r|r|r|}\hline 
field&$A_\m(x)$ & $\p_i(x)\&\pb_i(x)$ & $c(x)$ & $\cb(x)$ & $b(x)$ & $\f$ & 
$\o$ & $\s$ & $\sb$ & $\gb$ & $\g$\\ \hline 
dim&$1$&$3/2$&$0$&$2$&$2$&$0$&$0$&$4$&$4$&$2$&$2$\\ \hline 
$\f\Pi$&$0$&$0$&$1$&$-1$&$0$&$2$&$1$&$-2$&$-1$&$0$&$1$\\ \hline 
\end{tabular} 
 
\nobreak\vspace{.2cm}{\footnotesize 
{\bf Table 1.} Dimensions and ghost numbers of the fields.}  
\end{center} 

The constant ghost $\o$ generates global  gauge 
transformations of all the 
fields in the BRST algebra, except itself. The action does not depend
on it and one restricts observables to the equivariant cohomology     
$\S$, 
\eq 
\S =\{\OO   :\pad{\OO}{\o^a}=0; s\OO=0, \OO\neq s\FF\} 
\eqn{observables} 
where $\FF$ is itself $\o$-independent.  Since one is eventually  
interested only  in expectation values of gauge invariant functionals 
of $A,\pb $ and $\p$, the notion of {\it 
physical} observables can be further sharpened to functionals in 
the equivariant  cohomology with vanishing ghost number\cite{II,I}.

For the sake of completeness let me note that \equ{effaction} is the
most general covariant power counting renormalizable classical  action only for
an $SU(2)$ group. It depends on only two
gauge parameters $\a$ and $\d$. For $SU(n>2)$  one can still extend
the action by a BRST-exact and power counting renormalizable term
proportional to 
\eq
s\tr \cb A_\m(x) A_\m(x)= \tr b(x) A_\m(x) A_\m(x) -
\cb(x) A_\m(x) D^{A}_\mu c(x) -A_\m(x) \cb(x) D^{A}_\mu c(x)
\eqn{massterm}
The equation of motion of the Nakanishi-Lautrup field $b(x)$ is
then modified to
\eq 
b(x)=(\gb/g -\pa\ipr A(x))/\a -\d g[\cb(x), c(x)]-\rho g A^2(x)\ .  
\eqn{EMb1}
Upon elimination of $b(x)$ using\equ{EMb1} the most general power counting
renormalizable tree-level action of a covariantly gauge fixed
$SU(n>2)$ model depends on three gauge parameters $\a,\d$ and $\rho$
and has the form  
\eqa
S_1=S_0&+&2\rho\int_{\TT_D} dx\,\tr g (\pa\ipr A(x)) A^2(x) + 
g^2\a (\d-1) [\cb(x), c(x)]A^2(x) \nn\\
&&+ g\a \cb(x)\{A_\m(x), \pa_\m
c(x)\} +\half\rho g^2\a A^2(x) A^2(x) -\gb A^2(x)\ ,
\eqan{effaction1}
where $\{\cdot,\cdot\}$ denotes the anticommutator.
For $SU(n>2)$ the analysis of $II$ in this sense is incomplete and
should be repeated with  the more general action\equ{effaction1}. 
Let us only note here that a non-trivial value of the bosonic ghost
$\gb$ would simultaneously regulate the infrared behavior of the tree
level ghost- {\it and} gluon- propagators  if a stable  fixed point
with $\d\neq 0$ and $\rho\neq 0$ exists for the $SU(n>2)$ model. The
additional parameter $\rho$ could also soften
the constraints found in $II$ and lead to non-trivial fixed points of
the moduli- and parameter- space of an $SU(n>2)$ gauge theory also for
$n_f>n$. Since $\gb$ also couples to
$A^2$ in\equ{effaction1}, the analysis of $II$ becomes quite involved in
this more general setting.

My objective here is the relation between the perturbative
asymptotic scale $\Lambda_{\overline{MS}}$ and the fixed points of the moduli
space in  the simplest possible scenario and I will not embark on an
analysis of the rather complicated model described by\equ{effaction1}.
For the purist the following
considerations apply only to an $SU(2)$ gauge theory, since the tree
level action\equ{effaction} in this case is the most general one. 

I will in fact restrict myself to the RG-stable subset
of covariant gauges with $\d=1,\rho=0$. Using the result of $II$,
one can expect a nontrivial fixed point in the moduli space of this
restricted class of models for  $n_f=n$ in  $\a=3$ gauge. Choosing
$\d=1, \rho=0$ greatly facilitates the calculations, because  the
moduli $\f$ and $\s$ can be integrated out and there is no $\gb
A^2$-vertex in this class of gauges.  At $\d=1$ the coupling of $\phi$ to
$[\cb(x),\cb(x)]$ in\equ{effaction} vanishes, and the integration over
the $\phi$-moduli just sets $\s$ to zero. The quartic ghost coupling
also vanishes at $\d=1$, further simplifying the calculation.

For $\d=1, \rho=0$ the tree-level action\equ{effaction} thus
effectively is 
\be\label{effaction0} 
S=S_C + 2\int_{\TT_D} dx\ \tr&&\left[\frac{1}{2\a}(\pa\ipr A(x))^2 + 
\cb(x)\, D^A\ipr\pa c(x)  - [\cb(x),c(x)]\gb +\frac{1}{2\a g^2}\gb^2
\right. \nn\\  
&&\qquad \left. -\frac{1}{\a g}\gb \pa\ipr A(x) +\frac{1}{g} \g\cb(x) +
\frac{1}{g}\sb c(x) \right].  
\ee  
The perturbative expansion depends only on the moduli $\gb$ of
vanishing ghost number, since 
the moduli $\sb$ and $\g$ just serve to eliminate the (on a finite
torus normalizable) constant modes
of the ghost and anti-ghost.  
The action\equ{effaction0} is stable under renormalization and is the
starting point of this calculation. 

\section{The Measure on the Moduli Space}
The expectation value of an observable $\OO\in \Sigma$ is 
formally given by the path integral  
\eq 
\vev{\OO}=\NN \int d\gb  d\sb  d\g\  
\int\!\int [\DD \p\DD\pb\,\DD A\, \DD c\,\DD\cb]\ \OO\ \exp S,
\eqn{expect} 
which defines the perturbative loop expansion. It was shown in $I$
and $II$ that \equ{expect} is generally normalizable, i.e. that
$\vev{\one}\neq 0$ for an $SU(2)$ group. 
The global bosonic ghosts $\gb$ introduced by the 
equivariant  BRST-algebra are the only relevant moduli parameters of
the model\cite{II}. We indicated in\equ{expect} that the  
integration over this finite dimensional moduli-space should  
be performed {\it after} the path integral over    
dynamical fields. As far as the dynamical fields
are concerned, the moduli $\gb$ are thus parameters of the
action. On a torus, the integrations over
constant  ghosts $\sb$ and $\g$ can be performed explicitly. They
eliminate  the zero-momentum modes of the Faddeev-Popov 
ghosts. The bosonic ghost $\gb$ in principle can also be eliminated,
since $S$ depends on $\gb$ only quadratically. This would  result in a 
{\it nonlocal} 4-point interaction of the dynamical ghosts that gives
rise to infrared divergences in the loop expansion. A
resummation of the perturbation series is therefore necessary
and  the term $\tr\gb [\cb(x), c(x)]$ in\equ{effaction0} should be
treated as an unconventional mass term for the dynamical ghosts. The ghost
``masses'', $\gb$, are finally to be integrated with a certain
measure. The measure, $d\m(\gb)$, on the moduli space
\eq
d\m(\gb)=e^{\G(\gb)}\prod_{a=1}^{n^2-1} d\gb^a
\eqn{defG}
is described by the effective action $\G(\gb)$. We will calculate $\G(\gb)$ 
order by order in the loop-expansion for the dynamical fields. Note
that $\G(\gb)$ is proportional to  the volume of
Euclidean space-time to leading order in $L$. In the thermodynamic
limit the measure\equ{defG} therefore effectively constrains the
moduli space to the absolute maxima $\widetilde\gb$ of $\G(\gb)$. The
following calculation will determine this space.

\subsection{The Tree-level Measure}
To lowest order in $\hbar$ the measure on the moduli-space is Gaussian and
apparently  given by the quadratic term $\frac{L^D}{\a g^2}\tr\gb^2$ of
the action\equ{effaction0}. Note that the term
$\int \tr \gb \pa\ipr A$ 
in\equ{effaction0} vanishes on a torus with
periodic boundary conditions for the gauge field $A_\m(x)$. Since we 
eventually wish to consider the thermodynamic limit of the
symmetric torus, $1/L$ is not a very useful mass
scale. Introducing an arbitrary, but in the limit $L\rightarrow\infty$ 
finite,  renormalization mass $\m$,  the
tree-level effective action in $D=4-2\e$ dimensions can be  expressed in
terms of the  dimensionless quantities that also appear in the $1$-
and $2$-loop contributions
\eq
\Gamma^{tree}(\gb) =\frac{L^D\tr\gb^2}{\a
g^2}=-\frac{L^D}{\a g^2}\sum_{i=1}^n e^2_i =-(\m
L)^D\frac{16\pi^2}{n\a\gh^2}\sum_{1\leq i<j\leq n} v^2_{ij}\, , 
\eqn{treelevel}
where $\gh=g\m^{-\e}$ is the dimensionless coupling constant and the
$e_i,i=1,\dots,n$, are the real 
eigenvalues of the traceless  hermitian
matrix $i\gb=i\gb^a t^a$. Since $\sum_i e_i=0$, the effective action
is a function of the dimensionless differences 
\eq
v_{ij}:=\frac{e_i-e_j}{4\pi\m^2}\ ,
\eqn{defv}
of the eigenvalues only.

Note that\equ{treelevel} is the effective action in terms of  {\it
bare } parameters  $\a=Z_\a \a_R$, $\gh=Z_g \gh_R$  and $\gb=Z_\gb
\gb_R$. By calculating the (regularized) effective action to two
loops, I will determine the first few coefficients in the
loop-expansion of some of these renormalization constants in the
$MS$-scheme. The coupling $g$ and gauge parameter $\a$ are also 
related to vertices of the dynamical fields and one can thus check the
validity of Ward-identities to some extent explicitly. 

Before calculating loop corrections to the effective
  action note the following interesting fact. Let us for the moment
  ignore the constraints imposed by the 
  BRST-algebra\equ{sdef} and regard $\gb$ as the zero-momentum mode of
  a background field $\gb(x)$. The coupling $\gb(x)\pa\cdot A(x)$ of
  $\gb(x)$ to 
  the longitudinal gauge field in \equ{effaction0} in this case precisely
  compensates the   $\gb^2(x)$-term of  the tree-level effective action
  {\it for all 
  non-zero} momentum modes of $\gb(x)$.  The effective action for
  non-constant modes of $\gb(x)$ in fact would vanish to all orders in
the loop expansion due to a supersymmetric compensation between ghosts
and longitudinal gluons. This supersymmetric compensation is {\it
absent} for the constant part $\gb$ of $\gb(x)$. The effective action for
  constant $\gb$ does {\it not} vanish, and  the corresponding
  measure on the moduli space is therefore not flat. 
In this sense the non-trivial measure on the moduli space we will find
  is a consequence of the 
{\it lack} of supersymmetric compensation between ghost and
(unphysical) gluonic degrees of freedom with vanishing momentum. The
  same imbalance ensures that the partition function, proportional to
  the Witten index of the TQFT on the gauge group\cite{I,brs96}, does not
  vanish. This mechanism is reminiscent of the lack of supersymmetric
  compensation for the ground state in models with unbroken
  supersymmetry and thus perhaps worth noting. The following will determine
  the ground state configuration of the interacting theory. 

\subsection{The One-Loop Measure}
The measure on the moduli space is Gaussian to lowest order in $\hbar$
only. The dynamics depends on the value of $\gb$ and the  one-loop
radiative corrections to the effective action are depicted in
Fig.~1. These contributions to the effective action were
evaluated in $II$ (and for $SU(2)$ also in\cite{I}). For a $D=4-2\e$
dimensional torus, the leading contribution for $L\m\sim\infty$ is\cite{II} 
\eq
\Gamma^{1-loop}(\gb) = (\m L)^D \sum_{1\leq i<j\leq n} 2 v_{ij}^{2-\e}
\cos(\half\pi\e)\G(\e-2)\ .
\eqn{1loop}
\vskip .5cm
\hskip 5cm\psfig{figure=ghostloop1.ps,height=1.5in}
\nobreak\newline
{\small\baselineskip 5pt 
\noindent Fig.~1: Diagrammatic representation of 1-loop contributions
to the effective action of the moduli.} 

Defining
\eq 
t_{ij}:=\g_E-1 +\half\ln\left(v_{ij}^2\right)\ ,
\eqn{defs}
where $\g_E=0.577216\dots$ is Euler's constant, \equ{1loop} for
$\e\sim 0$ has the expansion 
\eq
\Gamma^{1-loop}(\gb) = (\m L)^D \sum_{1\leq i<j\leq n}
v^2_{ij}\left(\frac{1}{\e} +\half -t_{ij} +O(\e)\right)\,,
\eqn{1loope}
Note that $\G^{1-loop}$ does not depend on $\a$ nor on $\gh$. Since $\gb=Z_\gb
\gb_R$ with  $Z_\gb-1$ of $O(\gh_R^2/\e)$, $\Gamma^{1-loop}$  gives rise
to a term of order $\gh_R^2$ in the renormalized two-loop effective
action. This contribution is proportional to
\eqa
\left. \pad{}{s}\Gamma^{1-loop}(s\gb)\right|_{s=1} &=& (\m L)^D
\sum_{1\leq i<j\leq n} (-2)(v^2_{ij})^{1-\e/2}
\cos(\half\e\pi)\G(\e-1)\nn\\
&=&(\m L)^D \sum_{1\leq i<j\leq n} v^2_{ij}\left(\frac{2}{\e} - 2
t_{ij} +\e[1-\frac{\pi^2}{12} + t_{ij}^2] + O(\e^2)\right)
\eqan{der1loop}
where the term of $O(\e)$ still gives a finite contribution to the
renormalized two-loop effective action, since the coefficient $Z_\gb-1$ is
of $O(\gh_R^2/\e)$.

\subsection{The Two-Loop Measure}
The $\gb$-dependent part of the (unrenormalized) 2-loop  effective
action is depicted in Fig.~2. With the methods of $II$, the leading
contribution for $\m L\sim\infty$ can be computed for a $D$-dimensional torus. 
The zero-modes of the ghosts are absent in the mode expansion and
the zero-modes of the gluon field give a subleading contribution for
large $\m L$. For the leading contribution to the effective
action  the mode-sums can be replaced by integrals
since they are IR-finite in $D>2$ for $\gb\neq 0$. In appendix~A these
integrals are evaluated using conventional Feynman techniques with  the
result\equ{feynmanint}, valid for $2<D<4$. For $\e\sim 0$ the
expansion of \equ{feynmanint} has the form
\eqa
\Gamma^{2-loop}(\gb)\stackrel{{\scriptsize \e\sim 0}}{=}  (\m
L)^D\frac{n\gh^2}{32\pi^2}&&\sum_{1\leq i<j\leq n} 
v^2_{ij}\left\{\frac{3-\a}{2\e^2}
+\frac{2-(3-\a)[1/4+t_{ij}]}{\e}
+3-4t_{ij}\right.\nn\\
&&\qquad +
(\a-3)\{{\rm finite~terms}\} +O(\e)\Biggr\} \ .
\eqan{2loop}
\vskip .5cm
\hskip 5pt\psfig{figure=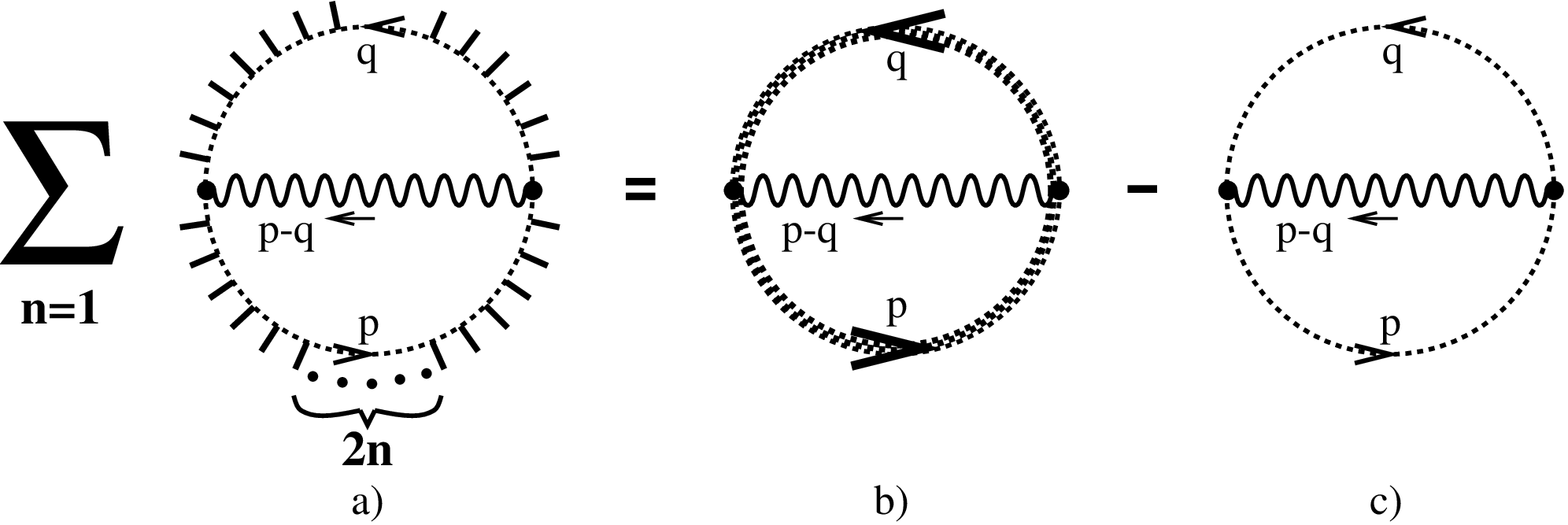,height=1.8in}
\nobreak\newline
{\small\baselineskip 5pt 
\noindent Fig.~2: a) Diagrammatic representation of the
$\gb$-dependent 2-loop contributions
to the effective action of the moduli. b) The contribution with the
tree-level ghost propagator (hatched) for fixed moduli $\gb\neq 0$. c)
The $\gb$-independent subtraction at $\gb=0$.} 

Note that the terms of\equ{2loop} proportional to $t_{ij}/\e$
{\it and} $1/\e^2$ vanish for $\a=3$. The two-loop contribution to the
effective action is therefore of order $\gh^2$ instead of order
$\gh^2 t_{ij}$ compared to the one-loop effective action\equ{1loope}
in this gauge. We will see that this suppression of the two loop
corrections in $\a=3$ gauge is not 
accidental  and that higher order terms of the form 
$\gh^{2n}s^{n+1}_{ij}$ with $n\ge 1$ are also absent in this gauge. 

\section{The Renormalized Two-Loop Effective Action}
We have so far computed the dependence of the effective action on the {\it
bare} variables $\gb,\gh$ and $\a$ in $D<4$ dimensions to two loops. To take
the limit $\e\rightarrow 0$, the effective action has to be
reexpressed in terms of renormalized quantities that remain finite in
this limit. We could of course have computed $\G$ as a function of
$\gb_R,\gh_R$ and $\a_R$ from the outset, by writing  
the action\equ{effaction0} in terms of renormalized fields and
couplings and including counterterm contributions in the computation
of the effective action. The two methods are entirely
equivalent, but the approach I chose requires less computational
effort. It avoids the computation of an infinite set of (counterterm)
diagrams which just give the derivative\equ{der1loop} of the infinite
set of 1-loop diagrams already computed.

To express the two-loop effective action by the renormalized
quantities, we simply substitute the expansions
\eqa
\gb&=&\gb_R Z_\gb=\gb_R (1+\frac{\gh_R^2}{16\pi^2} Z_\gb^{(1)} +
\frac{\gh_R^4}{(16\pi^2)^2} Z_\gb^{(2)} + O(\gh_R^6)) \nn\\
\gh^2 &=&\gh_R^2 Z_g^2=\gh_R^2 (1+\frac{\gh_R^2}{16\pi^2} Z^{(1)} +
\frac{\gh_R^4}{(16\pi^2)^2} Z^{(2)} + O(\gh_R^6)) \nn\\
\a&=&\a_R Z_\a=\a_R(1+\frac{\gh_R^2}{16\pi^2} Z_\a^{(1)} +
\frac{\gh_R^4}{(16\pi^2)^2} Z_\a^{(2)} + O(\gh_R^6))\, , 
\eqan{counter}
for the bare quantities in in the previous expressions. In the $MS$-scheme the
$Z^{(i)}_j$ are determined by requiring that they cancel only the
$1/\e^k$ terms. To the order we calculated,  
\eq
\Gamma^{2-loop}(\gb_R Z_\gb;\gh_R^2 Z_g^2, \a_R
Z_\a)=\Gamma^{2-loop}(\gb_R;\gh_R,\a_R) + O(\gh_R^4)\ ,
\eqn{2loopR}
and
\eq 
\Gamma^{1-loop}(\gb_R Z_\gb)=\Gamma^{1-loop}(\gb_R)
+Z_\gb^{(1)}\frac{\gh_R^2}{16\pi^2}
\left. \pad{}{s}\Gamma^{1-loop}(s\gb_R)\right|_{s=1}+ O(\gh_R^4)\ .
\eqn{1loopR}
Comparing \equ{2loop} and \equ{der1loop}, the $
v^2_{ij}t_{ij}/\e$ term in\equ{2loop} requires that we choose
\eq
Z_\gb^{(1)}= \frac{n(\a_R-3)}{4\e}
\eqn{zgb1}
in the $MS$-scheme. The remaining divergent terms of,
\eq
\Gamma^{1-loop}+\Gamma^{2-loop}= (\m L)^D \sum_{1\leq i<j\leq n}
v^2_{Rij}\left(\frac{1}{\e} +\frac{\gh^2_R n}{16\pi^2}
\left(\frac{(3-\a_R)}{4\e^2} +\frac{5+\a_R}{8\e}\right)\right)+{\rm
finite~terms}
\eqn{div12}
have to be canceled by the tree-level effective action\equ{treelevel}
expressed in terms of $\gb_R,\a_R$ and $\gh_R$
\eqa
\Gamma^{tree}(\gb_R Z_\gb; \gh_R^2 Z_g^2;\a_R Z_\a)&=&-(\m L)^D
\sum_{1\leq i<j\leq n}\frac{16\pi^2 v^2_{Rij}}{n\a_R\gh_R^2}
\left(1+\frac{\gh_R^2}{16\pi^2}(2
Z_\gb^{(1)}-Z^{(1)}-Z_\a^{(1)})\right. \nn\\
&&\hskip-12em\left.+\left(\frac{\gh_R^2}{16\pi^2}\right)^2 \left( (Z_\gb^{(1)} -Z^{(1)}
 -Z_\a^{(1)})^2 -Z^{(1)} Z_\a^{(1)} + 2 Z_\gb^{(2)} -Z^{(2)}
 -Z_\a^{(2)}\right)\right)+ O(\gh_R^4)   
\eqan{treeR}
Comparing with\equ{div12} one obtains two relations for the $Z's$ in
\equ{counter}: 
\eqa
Z^{(1)}+Z_\a^{(1)}&=&2 Z_\gb^{(1)}-\frac{n\a_R}{\e} =
-\frac{n(\a_R+3)}{2\e}\nn\\ 
2 Z_\gb^{(2)} &=&Z^{(2)}+ Z_\a^{(2)} + Z^{(1)} Z_\a^{(1)} -
\frac{n^2(13\a_R^2 +6\a_R +9)}{16\e^2} +\frac{n^2\a_R(5+\a_R)}{8\e} 
\eqan{relconst}
The first of these is a check of the two-loop
calculation, since  $Z^{(1)}$ and $Z_\a^{(1)}$ can also be determined
from the 1-loop radiative corrections to dynamical vertices. The
second relation in\equ{relconst} gives $Z^{(2)}_\gb$ in terms of
the renormalization constants for the couplings. With\cite{it80}
\eqa
Z^{(1)}&=&-\beta_0/\e= (\frac{2}{3}n_f-\frac{11}{3}n)/\e\nn\\
Z_\a^{(1)}&=&(\frac{13-3\a_R}{6}n-\frac{2}{3}n_f)/\e
\eqan{renorm1}
one verifies that the first relation in\equ{relconst} indeed holds. 

The terms proportional to $1/\e$ and $1/\e^2$ thus are seen to
vanish through  order $\gh_R^2$ when the effective action is expressed
in terms of $\gb_R,\gh_R$ and  $\a_R$. In the $MS$-scheme we have
\eqa
\G(\gb_R,\gh_R,\a_R)&=&(\m L)^D \sum_{1\leq i<j\leq n} v^2_{Rij}\left\{ -\frac{16\pi^2}{n\a_R\gh_R^2}   
+ 1/2 -t_{Rij}\right.\nn\\
&&\left. + \frac{n\gh_R^2}{16\pi^2}\left(3/2-2t_{Rij} + (\a_R-3)\{{\rm finite~terms}\}\right)\right\}
+O(\gh_R^4)  
\eqan{renormalized}
To simplify expressions allow me to sometimes abuse notation 
in the following and  denote renormalized quantities without an index
$(R)$ whenever confusion with bare quantities is unlikely.  

\subsection{Higher Loop Corrections  to the Effective Action}
Since  the coefficient $Z_\gb^{(1)}$, given by\equ{zgb1}, vanishes for
$\a_R=3$, the anomalous dimension of $\gb_R$ is of order $\gh_R^4$ in
this gauge. In $II$ it was observed that
$\a_R=3 $ is the stable non-trivial UV fixed point of the gauge parameter
$\a_R$ for $n_f=n$ because  $Z_\a^{(1)}$, given by\equ{renorm1}, then
vanishes as well. Using RG-arguments, it was shown\cite{II} that
$\a=3,\d=1,(\rho=0)$ is the CCG for $n_f=n$. Remarkably the leading
asymptotic correction
of order $\hbar$ in the effective action\equ{renormalized} is of the
form $\gh^2_R t_{Rij}$ rather than $\gh^2_R s^2_{Rij}$ for
$\a_R=3$. The two-loop contribution to the effective 
action in this gauge is thus suppressed by a factor $\gh^2$ compared to
the one-loop contribution. We will see explicitly that this is no
accident and that in fact all higher loop contributions are similarly
suppressed asymptotically in $\a_R=3$ gauge.

It is not difficult to show that a term of order $\gb^2_R t_{Rij}^k$ in the
renormalized effective action is associated with an 
unrenormalized $k$-loop contribution that diverges as $\gb^2/\e^k$. 
I will now argue that a generic 1-Particle Irreducible (1PI) $k\geq
2$-loop vacuum  diagram  that depends on $\gb$ at most diverges like
$\gb^2/\e^{k-1}$ for $\a=3$. 
The vacuum diagram we consider obviously must have at  least one ghost
loop and there must be at least two insertions of $\gb$ for it to
depend on $\gb$. We select a ghost loop with at least one insertion of
$\gb$.  For $n\geq 2$ the diagram is then  generically of the form shown in
Fig.~3, since a gluon must be emitted  somewhere before the insertion
of $\gb$ on the ghost loop and eventually must be reabsorbed by the
{\it same} ghost loop if the diagram is 1PI. 
\vskip .5cm
\hskip 4cm\psfig{figure=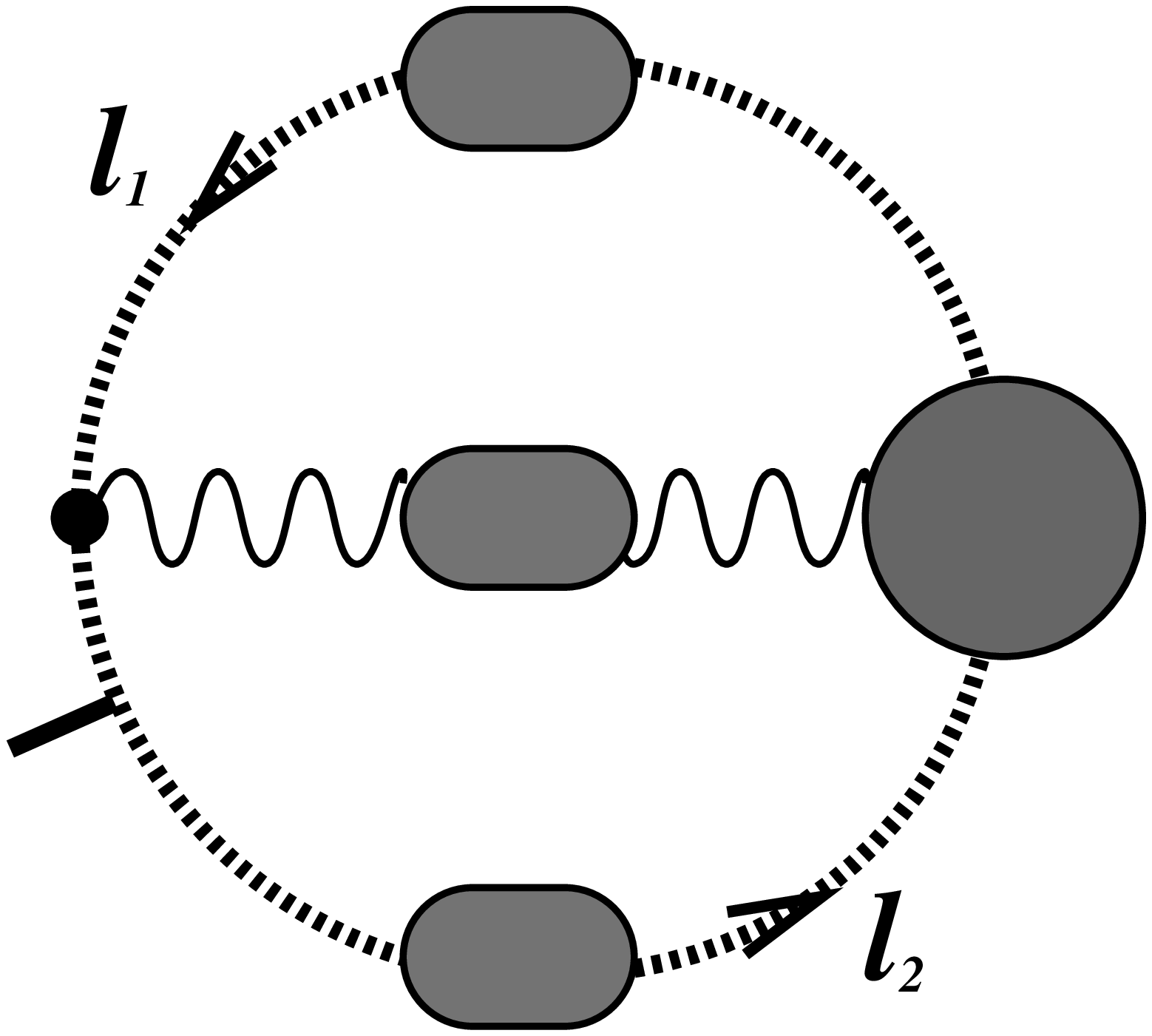,height=2.in}
\nobreak\newline
{\small\baselineskip 5pt 
\noindent Fig.~3: Diagrammatic representation of the sceleton of a
1PI contribution to the $\gb$-dependent effective action with two or
more loops.  Shaded areas denote generic parts of  a multi-loop
diagram that can be considered as a correction to the ghost-gluon
vertex or of the two-point functions themselves. See the text for details.} 

The
leading divergence of all the $k-2$ loop integrals apart from $l_1$
and $l_2$, is furthermore at most $1/\e^{k-2}$. The main point of the
argument is that such a divergence (if it occurs) cannot depend
on $\gb$ and is compensated by counterterms for the subdiagrams that
do not depend on the mass-scale $\gb$. The reason is that an insertion
of $\gb$ renders the ghost-gluon vertex and ghost selfenergy
superficially convergent (since one of the ghost momenta factorizes in
the gauge $\d=1$). The insertion of $\gb$ thus reduces the leading
divergence of these sub-diagrams by at least one power of $\e$. A
similar argument  holds for the gluon polarization, since a
single insertion of $(\hat\gb)_{ab}=f^{abc}\gb^c$ gives a vanishing
contribution due to Bose symmetry. As far as the leading divergence of
order $1/\e^{k-2}$ for the subdiagrams is concerned, the diagram of
Fig.~3 is thus equivalent to the two-loop diagram of Fig.~2 multiplied by
$(const./\e^{k-2})$. We have explicitly shown that the two-loop
contribution \equ{2loop} diverges 
only as $\gb^2/\e$ when $\a=3$. The leading divergence of the $k\geq 2$-loop
contribution is thus at most $\gb^2/\e^{k-1}$. The leading  dependence
on $\ln(\m^2)$, viz. $t_{Rij}$, 
of the renormalized  effective action in $D=4$ dimensions evaluated to
$N$-loops in $\a_R=3,\d=1$ gauge in the $MS$-scheme is therefore of
the form  
\eq
\G^{(N)}(\gb)=\frac{L^4}{16\pi^2} \sum_{1\leq i<j\leq n} (e_i-e_j)^2 
\left\{  -\frac{16\pi^2}{3n g^2} +\half-t_{ij}+ \sum_{k=1}^{N-1} c_k
\left(g^2 t_{ij}\right)^k\right\}\ , 
\eqn{leadinglog}
where the $c_k$ are $\m$-independent coefficients determined by the 
$(k+1)$-loop contribution, $e_i,i=1,\dots,n$ are the $n$ real
eigenvalues of the traceless hermitian matrix $i\gb_R$, and $g=g_{MS}$
is the renormalized coupling constant in the minimal subtraction ($MS$)
scheme. For $g^2 t_{ij}$ of $O(1)$ asymptotically, contributions to
the effective action from two and more loops are thus
suppressed relative to the one-loop contribution by at
least one order in $g^2$. The perturbative calculation of the
effective action in the limit $g\rightarrow 0$ thus gives a sensible
asymptotic expansion in the gauge $\a=3,\d=1,\rho=0$.    

\section{The RG-invariant maximum of the effective action} 
Since the effective action $\G(\gb)$ in\equ{defG} is essentially
proportional to the space-time volume for $(\m L)^4\sim\infty$, only absolute
maxima $\widetilde\gb$ of $\G(\gb)$ are of
relevance in the evaluation of expectation values\equ{expect} of
observables in the thermodynamic limit. We now determine the absolute
maxima $\widetilde\gb$ of $\G(\gb)$ in $\a=3,\d=1$ gauge. 

In general the eigenvalues of $\widetilde\gb$ will be
functions of the renormalization point $\m$, and will either vanish or
approach infinity as $\m\rightarrow\infty$. For $n_f=n$, they however approach a finite value proportional to $\Lambda_{\overline{MS}}$.
Maxima $\widetilde\gb$ of the effective  action \equ{leadinglog} for large
$\m^2/(\widetilde e_i-\widetilde e_j)$  satisfy 
\eqa
0&=&\left.\frac{8\pi^2}{L^4} \pad{}{s}\G^{(N)}(s\widetilde\gb;g,\m)\right|_{s=1}\nn\\
&=&\sum_{1\leq i<j\leq n} (\widetilde e_i-\widetilde e_j)^2
\left\{ -\frac{16\pi^2}{3n g^2}+\frac{c_1 g^2}{2} - \widetilde
t_{ij}\left[1-g^2\sum_{k=0}^{N-2} \widetilde
c_{k}\left(g^2 \widetilde t_{ij}\right)^k\right]\right\}\ ,  
\eqan{max}
where
\eq
\widetilde c_{k}(\m)= c_{k+1} +\half g^2 (k+2) c_{k+2}
\eqn{defctilde}
and $g^2$ is the running coupling constant of the $MS$-scheme. Its
dependence on the renormalization point $\m$ to two loops for
sufficiently large $\m/\L_{MS}$ is\cite{it80}  
\eq
\frac{g^2}{16\pi^2}=\frac{1}{\b_0\ln\m^2/\L_{MS}^2}
-\frac{\b_1}{\b_0}
\frac{\ln\ln\m^2/\L_{MS}^2}{(\b_0\ln\m^2/\L_{MS}^2)^2}\ ,
\eqn{running}
where
\eq
\b_0=\frac{11 n-2 n_f}{3} {\rm ~~and~~} \b_1=\frac{34 n^2-13 n_f + 3 n_f/n}{3}
\eqn{coeffs}
are the first two coefficients of the $\b$-function and $\Lambda_{MS}$
is the so defined  $\m$-independent asymptotic scale parameter of the
$MS$-scheme. For $n_f<\frac{11}{2}n$ quark flavors $g^2$ vanishes
logarithmically as $\m\rightarrow \infty$, implying 
asymptotic freedom. For \equ{max} to hold in the limit $\m\rightarrow
\infty$, $\widetilde t_{ij}$ must be of order $1/g^2$ and we see that
the sum in \equ{max} is a correction of order $g^2$ to $\widetilde
t_{ij}$. Since the perturbation series diverges for
$N\rightarrow\infty$, this is more precisely true in the sense of
an asymptotic expansion: for any fixed, but not
necessarily small, order in the loop expansion, the contributions from
two and more loops become negligible in 
the vicinity of the maximum $t_{ij}\sim\widetilde t_{ij}$ of the effective
action as $\m\rightarrow\infty$. 

The condition\equ{max} for the maximum of the effective action
can be
regarded as an asymptotic  definition of the coupling  
$g$ in terms of the renormalization point $\m$ and the scale
$\kappa^4=-\tr\widetilde\gb^2$. The situation is similar to 
comparing the asymptotic scales of two perturbative
renormalization schemes\cite{ce79}.  Using \equ{running} to eliminate  the
coupling $g^2$ in favor of the asymptotic scale $\L_{MS}$ of the
$MS$-scheme, \equ{max} in the limit $\m\rightarrow\infty$ implies that 
\eq
\frac{\displaystyle{3n \sum_{1\leq i<j\leq n} (\widetilde e_i-\widetilde
e_j)^2 \left(1-\g_E  
         -\half\ln\left(\frac{\widetilde e_i-\widetilde
e_j}{4\pi\m^2}\right)^2 \right)}}{\displaystyle{\sum_{1\leq i<j\leq n}
(\widetilde e_i-\widetilde e_j)^2}} =\b_0\ln\m^2/\L_{MS}^2  \, , 
\eqn{max1} 
Eq.\equ{max1} gives the asymptotic scale dependence of the
eigenvalues $\widetilde e_i(\m) $ corresponding to a maximum of the effective
action. We see that the  maximum approaches a finite
limit as $\m\rightarrow\infty$ only if $3n=\b_0$. From \equ{coeffs},
this is the case for 
\eq
n_f=n
\eqn{fixpointcond}
quark flavors.

For an  $SU(2)$ group and two quark flavors, \equ{max1} can be
explicitly solved for  the one independent eigenvalue $\widetilde e=\widetilde
e_1=-\widetilde e_2$ of $i\widetilde\gb$,
\eq
\widetilde e=\pm 2\pi e^{1 -\g_{E}}\L^2_{MS}=\pm\frac{e}{2}\L_{\overline{MS}}^2
\sim \pm(1.1658 \L_{\overline{MS}})^2 
\eqn{su2value}
We thus have that 
\eq
\kappa^4(n=n_f=2)=-\tr\widetilde\gb^2={\widetilde e}_1^2+{\widetilde
e}_2^2=\frac{e^2}{2}\L_{\overline{MS}}^4 
\sim (1.3864\L_{\overline{MS}})^4
\eqn{kappa2}

In the case of $SU(3)$ with three quark flavors, \equ{max1} by itself
is not sufficient to determine the {\it texture} of the eigenvalues of
$i\widetilde\gb$. However, if the maxima of\equ{leadinglog} are unique up
to permutations of the eigenvalues $\widetilde e_i$, the discrete symmetry
$\gb\rightarrow -\gb$ of the effective action must be equivalent to a
permutation of the eigenvalues $\widetilde e_i$. For an $SU(3)$ group this
requires that the absolute maxima of the effective action  are permutations of
\eq
\widetilde e_1=-\widetilde e_2=\widetilde e,\qquad \widetilde e_3=0\ .
\eqn{texture}
For $n_f=3$, \equ{max1} then determines the value of $\bar
e$ in terms of $\L_{\overline{MS}}$
\eq
\widetilde e=2^{-2/3} e (4\pi\L_{MS}^2 e^{-\g_E} ) =2^{-2/3} e
\L_{\overline{MS}}^2 \sim (1.3086 \L_{\overline{MS}})^2\ .
\eqn{su3value}
and the corresponding value for the scale
\eq
\kappa^4(n_f=n=3)=-\tr\widetilde\gb^2=2^{-1/3} e^2\L_{\overline{MS}}^4\sim
(1.5562\L_{\overline{MS}})^4
\eqn{kappa3}

For $SU(n>3)$ groups and $n_f=n$, symmetry arguments and
\equ{max1} alone are not sufficient to completely determine the
absolute maxima of the effective
action. The relation\equ{max1} however implies that one can find
the fixed points in the limit $\m\rightarrow \infty$ by (numerically) 
maximizing 
\eq
\G^{eff}(\gb;\L_{\overline{MS}})=-\frac{L^4}{32\pi^2} \sum_{1\leq i<j\leq
n} (e_i-e_j)^2 \ln \frac{(e_i-e_j)^2}{(\exp 3) \L_{\overline{MS}}^4}\ .
\eqn{Geff}
This is not to say that \equ{Geff} is the effective
potential for $\gb$. It cannot be, since the anomalous dimensions of
$\gb$ and $\a$ only vanish to order $\hbar$ for $\a=3$ and 
$n_f=n$. $\G^{eff}$ therefore does {\it not} satisfy the correct
renormalization group equation. We have
however seen that higher loop corrections become negligible in the {\it
vicinity} of maxima for $\m/\L_{\overline{MS}}\sim \infty$, since
\equ{max1} shows that they are of order $g^2$. $\G^{eff}$ 
correctly describes the variation of the effective action {\it near} its
maxima in the limit $g^2\rightarrow 0$ and thus correctly reproduces
\equ{max1} in this limit. Higher loop
corrections to\equ{leadinglog} in general are {\it not} negligible
sufficiently far from the maxima when  $\left|\ln|(e_i-e_j)/\m^2)|\right|\gg
1/g^2$. $\G^{eff}$ of \equ{Geff} in this case differs significantly from the
true effective action.
 
It was argued in $II$ that the
deviation of the true effective action from $\G^{eff}$ is of
no importance in the thermodynamic limit, since the measure
 on the moduli space, proportional to $e^{\G(\gb)}$, constrains
the moduli space of $\gb$ to the immediate vicinity of absolute
maxima of the effective action. Fluctuations of the moduli  vanish as
$1/L^2$. In the thermodynamic- ($L\rightarrow\infty$) {\it and} critical-
($g^2\rightarrow 0$) limit,  the measure on the moduli space is thus
effectively given by $\G^{eff}$. 

The following remark is perhaps of interest: our
perturbative evaluation is uncontrolled near $\gb=0$ and the
fact that $\G^{(N)}(0)$ and $\G^{eff}(0)$ vanish does not imply that the
true effective action also vanishes for $\gb=0$, since
$\left|\ln|(e_i-e_j)/\m^2|\right|\rightarrow \infty \gg 1/g^2 $ for {\it any 
finite} value of $\m$ in this case. Higher orders in the loop expansion are
therefore much more important than the computed ones and an asymptotic expansion of the effective action is pretty useless near $\gb=0$. One could
imagine that after a resummation of the perturbative expansion, the
factor $\gb^2$ of the effective action  cancels for $\gb\rightarrow 0$
and that the true effective action has a finite or perhaps even
divergent limit for $\gb\rightarrow 0$. 
In $II$ it was argued that an absolute  maximum of the true effective
action at $\gb=0$ would imply that nonperturbative effects dominate
even asymptotically, since the absolute maximum of the effective
action sets the asymptotic physical scale of the
model -- as we have seen, this maximum does not occur at $\gb=0$ in
any finite order of the loop expansion. This argument however does not
{\it a priori}
imply  that the true effective action {\it vanishes} at $\gb=0$, i.e. that
$\G^{eff}(\gb)$ not only reproduces the true effective action near its
absolute maximum, but also gives the difference in  energy density between
the broken and unbroken phases. The next section shows that
$\G^{eff}(\gb)$ does indeed reproduce the correct energy difference. 

\section{$\vev{\Theta_{\m\m}}$ in terms of $\L_{\overline{MS}}$}
We can also compute the energy density difference between the
broken phase with $\vev{\gb^2}=\widetilde\gb^2\ne 0$ and the
$SU(n)$-symmetric one with $\vev{\gb^2}=0$ by a change of scale, since
Euclidean invariance relates the vacuum expectation value of a
component of the energy
momentum tensor $\Theta_\mn$ to the vacuum expectation value of its
trace
\eq
\vev{\Theta_\mn}=\frac{\d_\mn}{4}\vev{\Theta_{\rho\rho}}
\eqn{vevT}
and an insertion of $\Theta_{\m\m}$ is the response to a change in
scale. In the chiral limit with $m_f=0$ the action\equ{effaction0} in
$D=4$ dimensions is invariant under an infinitesimal
dilation and a corresponding change of integration variables,   
\eqa
\d x &=&-x\,\d\l \qquad \Rightarrow  \d L=-L\, \d\l \nn\\
\d\Phi_i(x)&=&(d_i \Phi_i(x)-x_\m \pa_\m \Phi_i (x))\,\d\l
\eqan{dil}
where $\Phi_i$ is a generic  field variable and $d_i$ is
 its canonical dimension as given in Table~1. Note that the change of
 variables also includes the moduli, which are integration variables
 and in this sense  {\it differ} from mass parameters and coupling
 constants. The reason for the {\it formal} invariance of the massless
 theory  under a change of scale of course is that there are no 
 dimensionful parameters in  $D=4$. This is no longer true for the
 regularized model in
 $D<4$ (or for that matter any other regularization of the model) and
 \equ{dil} is not a symmetry of the renormalized theory (which does depend on a
 physical mass scale such as $\L_{\overline{MS}}$ that does not vanish in
 the chiral limit). One can compute the scale anomaly from the fact
 that  the (bare) coupling constants and fields
 of the dimensionally regularized model in $D<4$ effectively change
 when the fields are transformed according to\equ{dil}. This
 anomalous contribution to the trace of the energy momentum tensor of
 QCD was first derived by  Collins, Duncan and Joglekar\cite{co77}. They found
 that the dilation\equ{dil} is equivalent to a zero-momentum  insertion
 of the renormalized operator 
\eq 
\Theta_{\m\m}=\frac{\b(g)}{2g} F_\mn^a F_\mn^a +\sum_{f} m_f \bar\Psi_f\Psi_f
 (1+\g_m(g))
\eqn{traceEM} 
in gauge invariant correlation functions. In\equ{traceEM} $\b(g)$ is
the $\b$-function describing the scale dependence of the coupling
constant and $\g_m(g)$ is the anomalous dimension of the quark
masses. Naively one 
expects only an insertion of $\sum_f m_f\Psi_f\Psi_f$ in the massive
case. The rest of \equ{traceEM} is  the anomalous contribution to the
trace of the energy momentum tensor. The arguments of \cite{co77}
also hold for the action\equ{effaction0} and the trace anomaly need
not be recalculated here. It
suffices to note that the action\equ{effaction0} differs from the
model considered by\cite{co77} in the gauge fixing sector only. 
Since BRST-exact insertions in 
gauge invariant correlation functions vanish in any dimension, the net
effect of a dilation on gauge invariant observables is an insertion of
\equ{traceEM} also for our model. 
Taking the vacuum expectation value of\equ{traceEM} 
\eq
\vev{\Theta_{\m\m}}=\frac{\b(g)}{2g} \vev{F_\mn^a F_\mn^a} +\sum_{f} m_f
\vev{\bar\Psi_f\Psi_f}  (1+\g_m(g))
\eqn{vactraceEM}
we see that non-vanishing gluon and quark condensates imply a
reduction of the vacuum energy compared to a hypothetical scale
invariant phase in which these condensates  vanish.

Since a dilation\equ{dil} is equivalent  to an insertion of
$\Theta_{\m\m}$ at zero momentum, we can  compute
\equ{vactraceEM} from the effective action $\G(\gb)$ for the constant ghost
$\gb$  after all dynamical fields have been integrated out.  
The finite dimensional integral over the moduli $\gb$ itself does not
diverge and the  
limit $D\rightarrow 4$ of the (renormalized) effective action can be
taken {\it before} performing these integrals. The trace of the energy
momentum tensor is furthermore RG-invariant and therefore does not
depend on the renormalization scale $\m$. We can take the limit $\m\rightarrow
\infty$ to compute it. In this limit the  change of scale\equ{dil}
effectively results in  an insertion of $-L^4 n\tr\gb^2/(8\pi^2)$ since
$\G^{eff}(\gb)$ describes the measure  accurately in the vicinity of a
maximum of the effective action. Since the anomalous dimension of
$\gb$ and $\a$ vanish to 
order $g^2$ in $\a=3$ gauge, we obtain for $n_f=n$ flavors,
\eqa
L^4\vev{\Theta_{\m\m}}&=&\lim_{\m\rightarrow\infty}\vev{\left.(2s\pad{}{s}-4
L\pad{}{L})\G(s\gb;g(\m),\m)\right|_{s=1}}\nn\\
&=& \vev{\left.(2s\pad{}{s}-4
L\pad{}{L})\G^{eff}(s\gb;\L_{\overline{MS}})\right|_{s=1}}\nn\\
&=& -\vev{\frac{L^4}{8\pi^2} \sum_{1\leq i<j\leq n} (e_i- e_j)^2}\nn\\
&=& \frac{L^4 n}{8\pi^2} \tr\widetilde\gb^2 =-\frac{L^4
n}{8\pi^2}\kappa^4(n_f=n) 
\eqan{scaleing}

The expectation value of the trace of the energy momentum tensor can
thus be expressed in terms of the fixed point
$\widetilde\gb$ of the maximum of the effective action $\G(\gb)$. Since
the left hand side of\equ{scaleing} is the
expectation value of a gauge invariant operator, it is worth
emphasizing that the fixed point $\widetilde\gb$ of the effective action
is {\it not} gauge dependent although our {\it perturbative}
evaluation of $\widetilde\gb$ in terms of 
$\L_{\overline{MS}}$ for $n_f=n$ flavors apparently was restricted to the
gauge  $\a_R=3,\d=1,\rho=0$.  As emphasized in $II$,
$\a_R=3$ is however not just any particular gauge. It is the stable
fixed point of the gauge parameter for $n_f=n$. As shown in $II$ this
is related to the fact that $Z_\a=1+O(g^4)$ at $\a_R=3$ for
$n_f=n$. In this sense $\lim_{\m\rightarrow\infty} \a_R(\m)=3$ for {\it
whatever} gauge ($\a_R\neq 0$) one starts from  at finite $\m$. [To have
$\lim_{\m\rightarrow\infty} a_R(\m)\neq 3$ for $n_f=n$, the bare gauge
parameter $\a_B(\m)$ would have
to be $\m$-dependent -- invalidating the
RG-analysis. Note that Landau gauge, $\a_R=0$, is an unstable
fixed point for $n_f<\frac{13}{4}n$\cite{II}.] 
The ``dangerous'' leading logs in gauges $\a_R\neq 3$ are
related to the running of $\a_R$ and the anomalous dimension
of $\gb$ in these gauges. In the limit
$\m\rightarrow\infty$ in which $\a_R(\m)\rightarrow 3$, the contribution
to the fixed point of the effective action from ``dangerous logs'' cancels
(which can be inferred from the RG-equation). In a loop  expansion of the
effective action to {\it finite} 
order this cancellation is  not seen in gauges $\a_R(\m)\neq 3$
because  the perturbative expansion is analytic in $g^2$. Choosing the
fixed point $\a_R(\m)=3$  is {\it 
not } a choice of gauge in the {$\m\rightarrow\infty$} limit,
but rather presents a simplification of the asymptotic analysis. The value
$\a_R(\m)$ at finite $\m$ only determines how the fixed point $\a_R=3$
is approached\cite{II}.

The asymptotic evaluation of the expectation value of the trace of the
energy momentum tensor in terms of $\kappa$ for $n_f=n$ flavors in\equ{scaleing} thus does not depend on the choice of covariant
gauge at finite $\m$ (with the possible exception of Landau gauge,
$\a_R=0$, for which our asymptotic analysis is not valid).  We can
combine\equ{scaleing} and \equ{kappa3} to the
physically interesting relation 
\eq
\vev{\Theta_{\m\m}} =-\frac{3}{8\pi^2} \kappa^4(n_f=n=3)= -\frac{3 e^2}{4^{5/3}\pi^2}\L_{\overline{MS}}^4 \sim\, -\,\left(0.687\L_{\overline{MS}}\right)^4 
\eqn{tracelambda}
between the vacuum expectation value
of the trace of the energy momentum tensor and the asymptotic scale
parameter $\L_{\overline{MS}}$ of the modified minimal subtaction scheme of
an $SU(3)$ model with three quark flavors.

Using\equ{vevT} the difference $\Delta\e$ between the energy density
$\e_{\rm true}$ of the true ground state and that of the perturbative vacuum
with unbroken $SU(3)$-symmetry is 
\eq
\Delta\e=\e_{\rm true}-\e_{\rm
pert}=\frac{1}{4}\left(\vev{\Theta_{\m\m}}_{\widetilde\gb}
-\vev{\Theta_{\m\m}}_{\widetilde\gb=0}\right)=- \frac{3}{32\pi^2} \kappa^4
=-\frac{3e^2}{16^{4/3}\pi^2}\L_{\overline{MS}}^4\sim -(.486\L_{\overline{MS}})^4\ ,
\eqn{edensity}
for $SU(3)$ with three quark flavors. The difference\equ{edensity}
is precisely $\G^{eff}(\widetilde \gb)-\G^{eff}(0)$ of\equ{Geff}.

\section{Discussion and a comparison with phenomenology}
I wish to stress that the final result\equ{tracelambda} is a
determination rather than an estimate of the expectation value of the
trace of the energy momentum tensor for an $SU(3)$ gauge theory with
three quark flavors. It implies that the unbroken phase with global $SU(n)$
color symmetry, and thus $\widetilde\gb=\vev{\gb}=0$,  corresponds to
the  scale invariant phase with
$\vev{\bar\Psi_f\Psi_f}=0$ and $\vev{F_\mn^a F_\mn^a}=0$. The relation
\equ{tracelambda} furthermore shows that there are no {\it other}
non-perturbative contribution to the expectation value of the trace
of the energy momentum tensor. More precisely: if a {\it nontrivial}
fixed point $\widetilde\gb$ of the moduli $\gb$ exists, its magnitude {\it
gives} the expectation value of the trace of the energy momentum
tensor. As I have argued, this fixed point of the moduli space  in fact
does not depend on the gauge parameter 
$\a$. Eq.\equ{tracelambda} gives it an even broader, completely gauge
invariant, meaning. The contribution to the vacuum energy density from
non-perturbative field configurations such as instantons, monopoles
and whatever else there might be is subsumed by the expectation value
of the moduli  
$\gb$. The perturbative expansion around a particular ground state with
$\widetilde\gb\neq 0$ makes sense because Eq.\equ{tracelambda} implies
that this vacuum has the correct energy density. It is furthermore clear that
the modified perturbative expansion {\it automatically} includes
power corrections to physical correlation functions at large Euclidean
momenta. These power corrections first appear at the three-loop level
in correlation functions of gauge invariant quark currents. They arise due
to insertions of $\widetilde\gb$ in ghost loops. Due to global gauge
invariance of physical correlation functions, the power corrections
in the chiral limit  are of order $\widetilde \gb^2/p^4,\widetilde
\gb^3/p^6,\dots$etc. as one also  expects from the OPE\cite{wi69}.  
The normalization of the coefficients is, however, no
longer arbitrary. Eq.\equ{texture} and \equ{su3value} relate the
asymptotic expectation values of $\gb$ to the fundamental scale
$\L_{\overline{MS}}$ of the model. We furthermore know that
corrections to these asymptotic values are {\it analytic} in 
$g^2$ in the gauge $\a=3,\d=1,\rho=0$. They can thus be reliably
computed in the framework of perturbation theory. In this asymptotic
sense the $SU(3)$ model with $n_f=3$ flavors is solved, since
the non-analytic ``non-perturbative'' corrections of the OPE have been
determined in terms of the asymptotic scale parameter. One could
consider this an explicit realization of the program first proposed by
Stingl\cite{st96}. 

We have here obtained the fixed point of the moduli
space only for  the special case of an $SU(n)$-symmetric model with $n_f=n$
flavors. The analysis in $II$ suggests that a fixed point
$\widetilde\gb\neq 0$ can also be found for $n_f<n$. The perturbative 
analysis in this case is complicated by the fact that the CCG occurs
for $\d\neq 1$. The  moduli $\f$ and $\s$ then have to be taken into
account and cannot be trivially integrated out.  
A non-trivial fixed point of the moduli space of an
$SU(3)$-model with $n_f\sim 6>3$ quark flavors, if it exists, would be
of greater phenomenological interest. In the restricted space of
covariant gauges with $\r=0$ no such fixed point was
ascertained\cite{II}. We are currently investigating whether there is
a fixed point in the moduli space of the $SU(3)$-model for $n_f>3$ in
the enlarged set of covariant gauges with $\r\neq 0$.

By performing a two-loop calculation of the effective potential, we
have explicitly verified that {\it there is} a non-trivial
fixed point $\widetilde\gb$ in the moduli-space of an $SU(3)$-model with
$n_f=3$ quark flavors. The eigenvalues of $i\widetilde\gb$
have been related (see \equ{texture} and\equ{su3value}) to the
asymptotic scale parameter $\L_{\overline{MS}}$. The reader
probably  noticed that the one-loop effective action actually
suffices for this determination. The two-loop calculation in this
special case, however, explicitly demonstrates the suppression of higher order
corrections to the fixed point of the moduli space in CCG -- which
in\cite{II} was inferred from the RG-equation. The methods of
appendix~A can furthermore be used to calculate the loop expansion of
$SU(n)$-invariant (physical) correlators in terms of the
eigenvalues of the fixed point $i\widetilde\gb$. Although these
calculations are still somewhat cumbersome,  we can  in principle 
calculate {\it all} power corrections due to the moduli in this
model order by order in the modified perturbative expansion. 

A comparison with phenomenology is complicated by the fact that more
than three quark flavors  are dynamical at momentum scales above $\m^2\sim
m^2_c \sim (2 \GeV)^2$. An $SU(3)$-model with only three
flavors can thus only be expected to have reasonable accuracy at
energies where the effective coupling 
$\a_s$ is still rather large, $\a_s(2\GeV)\sim 0.35$. Let us
nevertheless attempt a comparison with the phenomenology of QCD
sum-rules while keeping this restriction in mind.

From moment sum rules in the $J/\Psi$ channel and chiral perturbation
theory, the LHS of \equ{tracelambda} can be 
estimated\cite{SVZ}. Any reasonable\cite{SVZ,PDGq,Na97} value for the
gluon condensate, $\vev{\bar\Psi\Psi}$ and the light quark masses
gives that the quark contribution to $\vev{\Theta_{\m\m}}$ is at most
$20\%$. It is thus comparable with the systematic error in the determination of
the gluon condensate. For a rough comparison with QCD-sumrules (and that is
unfortunately all we can hope for), \equ{tracelambda} essentially
states that
\eq
\left(\frac{2}{3} \L^{(3)}_{\overline{MS}}\right)^4\sim\frac{8}{9}\vev{\Theta_{\m\m}}\sim
\vev{\frac{\a_s}{\pi} F^a_\mn F^a_\mn}
\eqn{estimate1}
where $\L^{(3)}_{\bar {MS}}$ is the asymptotic scale
parameter of the modified minimal subtraction scheme for a
three-flavor model. The best determination of the gluon condensate at
the highest energies for which only three flavors are dynamical 
is obtained using QCD-sumrules for the  charmonium system. These balance
perturbative contributions against power corrections. Using ratios of
moments, Shifman, Vainshtein and Zakharov (SVZ)\cite{SVZ} originally
estimated 
\eq
\vev{\frac{\a_s}{\pi} F^a_\mn F^a_\mn}_{SVZ}\sim (330\pm 30 \MeV)^4\quad{\rm
with}\quad \a_s(2.5\GeV)=0.2 \quad{\rm and}\quad m_c\sim 1.26 \GeV
\eqn{svzval}
Most recent high-energy data indicate that their value for $\a_s$ is a
bit low and that $\a_s(2.5\GeV)\sim 0.3$ would be
more appropriate (see Fig.~4). Since $\a_s$ determines the perturbative
contribution to the moments, the extracted condensate value 
depends somewhat on $\a_s$. In the method used
by\cite{SVZ} it is also relatively sensitive to $m_c$.
\vskip .5cm
\hskip 2.cm\psfig{figure=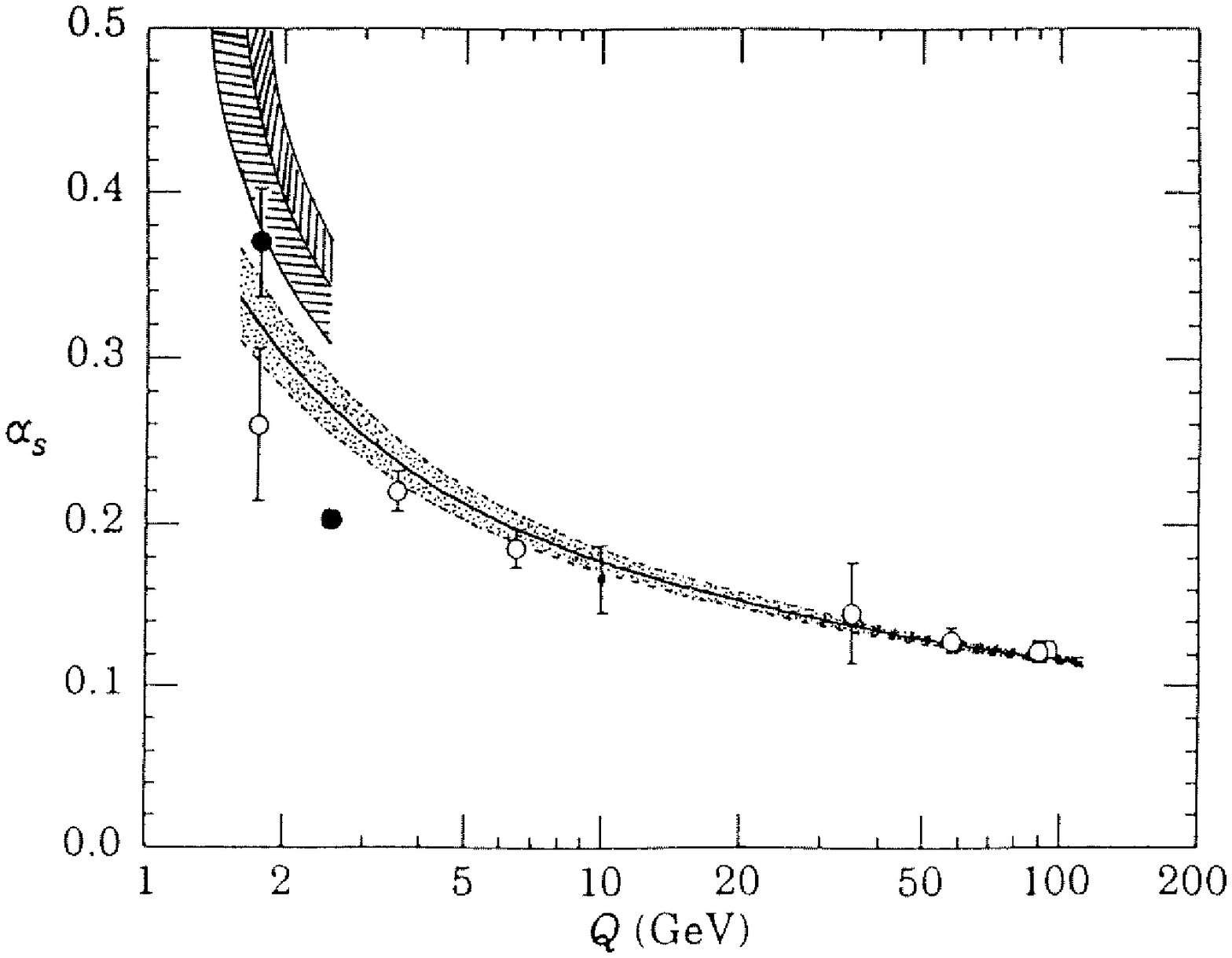,height=4.in}
\nobreak\newline
{\small\baselineskip 5pt 
\noindent Fig.~4: The running strong coupling constant $\a_s$ of the
modified minimal subtraction (${\overline{MS}}$) scheme. The
experimental data and best fit (solid line) with its 
$1\sigma$ error (grainy area) are reproduced from
figure~9.2 of the summary by the Particle Data Group\cite{PDG}. I have
highlighted the experimental datapoint from $\tau$-decay at $\sim
1.8\GeV$ in solid and also included with a solid point the value of
$\a_s(2.5\GeV)\sim 0.2$ used in the determination of the gluon
condensate by\cite{SVZ}. Also shown are two bands derived from
estimates of the gluon condensate. The upper band corresponds to
Narison's\cite{Na97}, whereas the 
lower band is obtained from Shifman, Vainshtein and
Zakharov's\cite{SVZ} value for the gluon 
condensate. Using\equ{estimate1} these estimates imply
$\L^{(3)}_{\overline{MS}}=580\pm 20\MeV$, respectively 
$\L^{(3)}_{\overline{MS}}=500\pm 40\MeV$. To appreciate the
quality of the fit note that data above $Q\sim 20\GeV$,
for which power corrections are negligible, are compatible with a
somewhat larger value of $\L^{(5)}_{\overline{MS}}$.} 
  
A recent analysis by Narison using double ratios of
moments\cite{Na97} reaches the conclusion that 
\eq
\vev{\frac{\a_s}{\pi} F^a_\mn F^a_\mn}_{Narison} \sim (375-400
\MeV)^4\quad{\rm with}\quad \a_s(1.3\GeV)=0.64^{+0.36}_{-0.18}
\eqn{Narvalue}
The advantage of this method compared to the previous one is that the
double ratios of moments do not depend on the
charm mass $m_c$ in leading order. The estimate\equ{Narvalue} for the
gluon condensate is about twice that of\cite{SVZ}.  The strong
coupling constant $\a_s$ in this analysis was obtained at the
optimal scale for  the  
$\chi_c(P_1^1)-\chi_c(P_1^3)$ mass-splitting. If a
perturbative evolution from $\m\sim 1.3\GeV$ to $\m\sim M_Z$ is
justified, the value for the coupling in\equ{Narvalue} 
corresponds to\cite{Na97} $\a_s(M_z)\sim 0.127\pm 0.011$ (see also
Fig.4).    

With\equ{estimate1} the SVZ-value\equ{svzval} for the gluon condensate
gives $\L^{(3)}_{\overline{MS}}\sim 500\pm 40 \MeV$. Using the three-loop
evolution\cite{PDG} of the corresponding coupling constant  (for
$n_f=n=3$) one obtains $\a_s(2.5 \GeV)\sim 0.325\pm 0.015$. Although 
compatible with other recent determinations of  $\a_s(2.5\GeV)$
(see Fig.4),  this value is {\it not} consistent with
$\a_s(2.5\GeV)\sim 0.2$ that was used for the sum-rule estimate of the
gluon condensate. 

The more recent determination\equ{Narvalue} for the gluon condensate
with\equ{estimate1} corresponds to $\L^{(3)}_{\overline{MS}}\sim 580\pm
20\MeV$. Again using the three-loop evolution\cite{PDG} this
implies $\a_s(1.3\GeV)=0.70\pm 0.04$, or equivalently,
$\a_s(2.5\GeV)=0.356\pm 0.008$. While consistent with the coupling
that was used\cite{Na97} to extract the 
gluon condensate, this value for $\a_s(2.5\GeV)$ is somewhat large 
compared to other determinations of $\a_s$ (see Fig.~4).  
Perhaps one should note in this context that the second term of the
$\beta$-function is almost $40\%$ of the first at
$\a_s=0.7$. Radiative corrections to the gluon condensate itself (which after
all give $\beta(g)/2g$ as the coefficient of $
\vev{F^2}$) and  higher loop  corrections on the perturbative side of
the sum-rules do not appear to be entirely negligible at these large
values for $\a_s$. The quoted  
error of only $15\%$\cite{Na97} for the gluon condensate
is therefore perhaps not a very conservative estimate of these systematic
errors. 

Using \equ{estimate1} and the three-loop evolution\cite{PDG} for
$\a_s$, one can also proceed in the
opposite direction and estimate $\L^{(3)}_{\overline{MS}}$ and thus the gluon
condensate from  $\a_s$ at scales
$\m<2.5\GeV$. This determination of the gluon condensate is only 
as accurate as the estimate for the asymptotic scale, which suffers from
the logarithmic relation between the latter and $\a_s$.
Probably one of the best experimental determinations of $\a_s$ at
relatively  low energies  is the one  from $\tau$-decay. The
central value of $\a_s(m_\tau)=0.370\pm 0.033 $
includes an estimate of hadronic power corrections to the
decay. Neglecting these would, however, change $\a_s(m_\tau)$ by less than the
quoted error\cite{PDG}. 

$\a_s(m_\tau)=0.370\pm 0.033$  corresponds to
$\L^{(3)}_{\overline{MS}}=450\pm80\MeV$ and with\equ{estimate1} implies a
gluon condensate of $\vev{\frac{\a_s}{\pi} F^2}\sim 300\pm 60\MeV$. Although
only a very rough value for the gluon condensate, it agrees
within errors with the one obtained by Shifman, Vainstein and Zacharov
(but for a considerably larger coupling). It is also compatible with Narison's
lowest estimate.  As Fig.~4 shows, this state of affairs
is not conclusive, but I consider it encouraging in
view of the fact  that the errors on $\a_s$ and  $\vev{\frac{\a_s}{\pi}
F^2}$ are almost wholly systematic.

In summary: the {\it exact} relation\equ{tracelambda} between the asymptotic
scale $\L_{\overline{MS}}$ of the modified minimal subtraction scheme
and $\vev{\Theta_{\m\m}}$ for an $SU(3)$ gauge
theory with three quark flavors is in reasonable agreement with
current estimates of the gluon-condensate and $\a_s$.

Note finally that $\L^{(3)}_{\overline{MS}}\sim 500\MeV$ (compatible
with Shifman, Vainstein and Zakharov's value for the gluon
condensate and $\a_s$ from $\tau$-decay) in\equ{edensity} corresponds
to a vacuum energy density of about $-455\MeV/\fm^3$. Assuming that  the
$SU(3)$-symmetry is essentially restored within a sphere of radius
$\sim 0.8\fm$ inside a nucleon, one naively obtains $M\sim
970\MeV$ for the mass of such a ``bubble''. 

\noindent ACKNOWLEDGMENTS\newline
I would like to thank L.~Spruch and D.~Zwanziger for their invaluable
support. 

\section{Appendix A: Calculation of the two-loop contribution to the
effective action}  

In the gauge $\d=1,\r=0$ the only two-loop contribution to the
effective action is the one shown graphically in Fig.~2. For  a
sufficiently large symmetric $D=4-2\e$ dimensional torus Fig.~2 is the
diagrammatic representation for the expression
\eqa
\Gamma^{2-loop}(\gb)&=& \frac{\gh^2}{2}\m^4 (\m L)^D\gh^2\int \frac{d^D p\, d^D
q}{(2\pi\m)^{2D}}\, p_\m q_\n D^{ab}_\mn(p-q) f^{acd} f^{bef}\times\nn\\
&&\qquad\left(D^{cf}(p;\gb) D^{ed}(q;\gb) -D^{cf}(p;0) D^{ed}(q;0)\right) 
\eqan{2loop1}
where $\gh=g \m^{-\e}$ is the dimensionless coupling constant and $L$
is the linear dimension  of the symmetric torus. In\equ{2loop1} $D_\mn^{ab}(k)$ is the tree level gluon propagator
\eq
D_\mn^{ab}(k)=\d^{ab}(\d_\mn k^2+(\a-1)k_\m k_\n)/k^4
\eqn{gluprop}
and $D^{ab}(k;\gb)$ denotes the tree-level ghost-antighost
correlator for fixed moduli $\gb$. Defining  the matrix
$(\hat\gb)_{ab}=f^{abc}\gb^c$ and using the integral
parametrization
\eq
D^{ab}(k;\gb)=(\d^{ab} k^2 +f^{abc}\gb^c)^{-1}=\int_0^\infty d\l
e^{-\l k^2} \left(e^{-\l\hat\gb}\right)_{ab}\ ,
\eqn{ghostprop}  
for the ghost propagators, the $D$-dimensional momentum integrals in
\equ{2loop1} can be separated from the color-summations. 
The expression for $\Gamma^{2-loop}$  becomes,
\eqa
&&\Gamma^{2-loop}(\gb)= \frac{\gh^2}{2}\m^4 (\m L)^D\int_0^\infty
d\l_1\int_0^\infty d\l_2
\left\{f^{dac}\left(e^{-\l_1\hat\gb}\right)_{cf}f^{fae}
\left(e^{-\l_2\hat\gb}\right)_{ed} - f^{dac}f^{cad}\right\}\nn\\  
&&\quad\times\int \frac{d^D p\, d^D q}{(2\pi\m)^{2D} (p-q)^4} \left((p-q)^2 pq +(\a-1)(p^2-
pq)(pq-q^2) \right) e^{-\l_1 p^2-\l_2 q^2}
\eqan{2loop2}
The  momentum space integrals can be performed in $D>2$ dimensions 
\eqa
&&{\hskip-2em}\int_{}^{} \frac{d^D p\, d^D q}{(2\pi\m)^{2D} (p-q)^4}
\left((p-q)^2 
pq +(\a-1)(p^2- 
pq)(pq-q^2) \right) e^{-\l_1 p^2-\l_2 q^2} =\int_0^\infty \frac{\l
d\l}{2} \times\nn\\
&&{\hskip-2em}
\left\{ \pad{^2}{\l_1\pa\l} +\pad{^2}{\l_2\pa\l}
+\frac{\a-1}{2}\left(\pad{}{\l_1}-\pad{}{\l_2}\right)^2-
\frac{\a+1}{2}\pad{^2}{\l^2}\right\}\int_{}^{}\frac{d^D p\, d^D
q}{(2\pi\m)^{2D}}  e^{-\l_1
p^2-\l_2 q^2-\l (p-q)^2}\nn\\
&&=\frac{D-1-\a(D-3)}{D-2}\frac{\l_1 \l_2}{(4\pi\m^2)^D (\l_1\l_2)^{D/2}
(\l_1+\l_2)^2}
\eqan{momint}  

I next evaluate the color trace in\equ{2loop2} in terms of the
(real) eigenvalues $e_i,\,i=1,\dots,n$ of the hermitian matrix
$i\gb=it^a\gb^a$. Note that
\eq
\left(e^{-\l\hat\gb}\right)_{ab}=-2\tr t^a e^{\l\gb} t^b e^{-\l\gb}
\eqn{fundamental}
with the  convention that  the anti-hermitian generators $t^a$ of the
fundamental representation of $SU(n)$ are normalized to $\tr t^a
t^b=-\half\d^{ab}$. Repeatedly using the completeness relation
\eq
t^a_{ij} t^a_{kl}=-\half(\d_{il}\d_{kj}-{1\over n}\d_{ij}\d_{kl})
\eqn{complete}
of the generators in the fundamental representation of $su(n)$, the
color trace in\equ{2loop2} can be rewritten as follows:
\eqa
&&{\hskip-2em}f^{dac}\left(e^{-\l_1\hat\gb}\right)_{cf}f^{fae}
\left(e^{-\l_2\hat\gb}\right)_{ed} = 4\tr [t^d,t^a] e^{\l_1\gb} t^f e^{-\l_1\gb}\tr [t^f,t^a] e^{\l_2\gb}
t^d e^{-\l_2\gb}\nn\\  
&&=-2\tr[ e^{\l_1\gb} t^f e^{-\l_1\gb}, t^d] [ e^{\l_2\gb}t^d
e^{-\l_2\gb}, t^f] \nn\\
&&=\tr e^{\l_2\gb}\tr e^{\l_1\gb} t^f e^{-(\l_1+\l_2)\gb} t^f + \tr
e^{-\l_2\gb}\tr e^{-\l_1\gb} t^f e^{(\l_1+\l_2)\gb} t^f - 2\tr
e^{-\l_2\gb} t^f\tr e^{\l_2\gb} t^f \nn\\
&&=\tr\one -\half\tr e^{\l_2\gb}\tr e^{\l_1\gb}\tr e^{-(\l_1+\l_2)\gb}
-\half \tr e^{-\l_2\gb}\tr e^{-\l_1\gb}\tr e^{(\l_1+\l_2)\gb}\nn\\
&&= -{\sum_{ijk}}^\prime \cos(\l_1(e_i-e_k)+\l_2(e_j-e_k))\ ,
\eqan{trace}
where the symbol ${\sum_{ijk}}^\prime$
denotes the sum over all triplets $(i,j,k)$ except those with $i=j=k$. The
first term in the expansion of the cosine does not 
depend on the eigenvalues and corresponds to the diagram shown in
Fig.2c. This $\gb$-independent term of the effective action can be
absorbed in the normalization $\NN$ of\equ{expect} and has been
subtracted in\equ{2loop1} since it is of no
physical interest. Using \equ{trace}, the color trace in\equ{2loop2}
expressed by the eigenvalues of $i\gb$ is,
\eqa
f^{dac}\left(e^{-\l_1\hat\gb}\right)_{cf}f^{fae}
\left(e^{-\l_2\hat\gb}\right)_{ed} - f^{dac} f^{cad} &=& {\sum_{ijk}}^\prime
(1-\cos(\l_1(e_i-e_k)+\l_2(e_j-e_k)))\nn\\ &&{\hskip-3em} =2{\sum_{ijk}}^\prime
\sin^2((\l_1(e_i-e_k)+\l_2(e_j-e_k))/2)\ .
\eqan{trace1}
When none of the eigenvalues $e_i, i=1,\dots,n$ coincide,
the argument of the sine does not vanish generically for arbitrary
values of $\l_1$ and $\l_2$ and regulates the infrared behaviour of
the integrand in\equ{2loop2}. Inserting\equ{trace1} and\equ{momint}
in\equ{2loop2}, the parametric expression for $\Gamma^{2-loop}$ is
\eqa 
\Gamma^{2-loop}(\gb)&=&\gh^2 (\m
L)^D\frac{D-1-\a(D-3)}{16\pi^2(D-2)}{\sum_{ijk}}^\prime  
\int_0^\infty \frac{\l_1 d\l_1}{\l_1^{D/2}}\int_0^\infty
\frac{\l_2 d\l_2}{\l_2^{D/2}}
\frac{\sin^2(\half(\l_1 v_{ik}+\l_2 v_{jk}))}{(\l_1 +\l_2)^2}\nn\\
&=&\gh^2(\m L)^D\frac{D-1-\a(D-3)}{16\pi^2(D-2)}{\sum_{ijk}}^\prime 
\int_0^1 dx\int_0^\infty
d\l \frac{\sin^2(\half\l(x v_{ik}+(1-x)
v_{jk}))}{\l^{D-1} (x(1-x))^{(D-2)/2}}\nn\\
&&
\eqan{2loop3}
where $v_{lm}$ is the dimensionless difference of eigenvalues defined
in\equ{defv}. Feynman's parametrization for the integrals shows that the
overall degree of divergence of the graph is logarithmic, as could
have been expected.  The integration over $\l$ in\equ{2loop3} is readily
performed in $D<4$ dimensions and we are 
left with an integral over a single Feynman parameter
\eqa
\Gamma^{2-loop}(\gb)&=&\gh^2(\m L)^D\frac{D-1-\a(D-3)}{16\pi^2(D-2)}
\cos\left(\frac{D\pi}{2}\right)\Gamma(2-D)\nn\\
&&\qquad\times {\sum_{ijk}}^\prime
\int_0^{1/2} dx \left(\frac{ (x v_{ik} 
+(1-x) v_{jk})^2}{(x(1-x))}\right)^{(D-2)/2}\,,
\eqan{feynmanint}  
where  the symmetry $i\leftrightarrow j$ under the
sum was used. Changing the integration variable to $y=x/(1-x)$ and
introducing $\e=(4-D)/2$, the integral in \equ{feynmanint} can also be written 
\eq
\int_0^{1/2} dx \left(\frac{ (x v_{ik} 
+(1-x) v_{jk})^2}{x(1-x)}\right)^{(D-2)/2}=\int_0^1
\frac{y^{1-\e} dy}{(1+y)^2} \left[\left(v_{ik}
+\frac{v_{jk}}{y}\right)^2\right]^{1-\e}
\eqn{feynmanint1}
For $\e\rightarrow 0_+$ \equ{feynmanint1} diverges as $v^2_{jk}/\e$. To
isolate this divergence of \equ{feynmanint1},
consider 
\eq
\int_0^1 \frac{dy}{(1+y)^2} \left(\frac{v^2_{jk}}{y}\right)^{1-\e}=
v^2_{jk}\left( \frac{1}{\e} -\half-\ln(2v^2_{jk})]
+\frac{\e}{2} [\frac{\pi^2}{6} 
+\ln(4)+(\ln(v^2_{jk}))\ln(4e v^2_{jk})]\right.+\dots
\eqn{divergence}
and define the for
$\e>-1$ and arbitrary real values of $a$ and $b$ regular integral  
\eqa
I_\e(a,b)&:=&\int_0^1 \frac{dy}{(1+y)^2}
\left\{(y a^2 
+b^2/y +2 a b)^{1-\e}-(b^2/y)^{1-\e}\right\}\nn\\
&=& \int_0^1 \frac{dy}{(1+y)^2}(y a^2 +2 a b)\nn\\
&&-\e\int_0^1 \frac{dy}{(1+y)^2}\left\{
(y a^2 +2 a b)\ln\left(y\left(a +\frac{b}{y}\right)^2\right) +
\frac{b^2}{y}\ln\left(\frac{b + y a}{b}\right)^2\right\} + O(\e^2)\nn\\
&=& a b +(\ln 2 -\half) a^2 + O(\e)
\eqan{finite}
In gauges $\a\neq 3$ the term of order $\e$ in\equ{finite} 
gives  a finite contribution to $\Gamma^{2-loop}$ in $D=4$ and would
have to be computed. For $\a=3$ -- the critical gauge that will
ultimately  interest us -- the 
 two-loop effective action simplifies greatly: the factor $D-1
-\a(D-3)$ in\equ{feynmanint} in this gauge is of order $\e$. The term
of order $\e$ in\equ{finite} thus 
gives a finite contribution to the 4-dimensional effective action
proportional to $(\a-3)$. This (finite) contribution will be of no further
interest and  for the sake of brevity will not be given in the form of
more elementary integrals
\footnote{algebraic programs such as MATHEMATICA readily express
the term proportional to $\e$ in\equ{finite} in the form of 
dilogarithms and products of logs.}.     

Making use of the symmetric summation over  eigenvalues,
\equ{finite} and \equ{divergence} together lead to,
\eqa
{\sum_{ijk}}^\prime && \int_0^1 \frac{y^{1-\e} dy}{(1+y)^2}\left[\left(v_{ik}
+\frac{v_{jk}}{y}\right)^2\right]^{1-\e}={\sum_{ijk}}^\prime
I_\e(v_{ik},v_{jk})+ \int_0^1 \frac{dy}{(1+y)^2} (v^2_{jk}/y)^{1-\e} \nn\\
&=&2n\sum_{1\leq i<j\leq n} v^2_{ij}\left
( \frac{1}{\e} -\half -\ln(v^2_{ij}) +O(\e)\right)
\eqan{expeps}

Expanding\equ{feynmanint} for $\e=(4-D)/2\rightarrow 0_+$ 
using\equ{expeps}, the 2-loop
contribution to the effective action  for a symmetric torus with volume
$(\m L)^D\sim\infty$ gives \equ{2loop} of the main text.

}
\begin{thebibliography}{99} 
\bibitem{wi69} K.G.Wilson, \pr{179}{69}{1499} for a review of the
application to QCD see  S.Narison, {\it QCD Spectral Sum Rules} (World
  Scientific, New   Jersey 1989);
\bibitem{II} M.Schaden, A.Rozenberg, \pr{D57}{98}{xxx}
\bibitem{I} L.Baulieu and M.Schaden, Preprint hep-th/9601039, to
  appear in IJMPA;
\bibitem{brs96} L.Baulieu, A.Rozenberg, M.Schaden,\pr{D54}{96}{7825};
\bibitem{tH79} G.'t Hooft, \np{B153}{79}{141} {\em ibid},
  \cmp{81}{81}{267} {\em ibid}, {\em Acta 
  Phys. Austriaca Suppl. XXII} (1980) 531; {\em ibid}, {\em
  Phys. Scr.}24 (1981) 841;
\bibitem{Hosotani} Y.~Hosotani, \pl{226B}{83}{309} D.~Toms, \pl{226B}{83}
{445} J.~E.~Hetrick and C.-L.~Ho, \pr{D40}{89}{4085} A.~McLachlan, \np{B338}
{90}{188}
\bibitem{ve79} G.Veneziano,\np{B159}{79}{213}
\bibitem{it80} C.Itzykson and J.--B.Zuber, {\it Quantum Field Theory}, 
               McGraw--Hill (1980);
\bibitem{ce79} W.Celmaster and R.J.Gonsalves, \pr{D20}{79}{1420}
\bibitem{co77} J.C. Collins, A. Duncan, S.D. Joglekar, \pr{D16}{77}{438}
\bibitem{st96} M. Stingl, \journal{\em Z. Phys.}{A353}{96}{423} and
references therein;
\bibitem{SVZ} M.A. Shifman, A.I. Vainshtein, V.I. Zakharov,
\np{B147}{79}{448}
\bibitem{PDGq} Particle Data Group, \pr{D54}{96}{303ff}
\bibitem{Na97} S. Narison, \journal{\em
Nucl. Phys. Proc. Suppl.}{54A}{97}{238}
\bibitem{PDG} Particle Data Group, \pr{D54}{96}{77ff}
\end{thebibliography}
\end{document}